\documentclass[sigconf]{acmart}
\AtBeginDocument{%
  \providecommand\BibTeX{{%
    \normalfont B\kern-0.5em{\scshape i\kern-0.25em b}\kern-0.8em\TeX}}}

\setcopyright{acmcopyright}
\copyrightyear{2021}
\acmYear{2021}
\acmDOI{10.XXXX/XXXXXXX.XXXXXXX}

\usepackage[utf8]{inputenc}
\usepackage{array, multirow}
\usepackage{soul}
\usepackage{pifont}
\usepackage{amsmath} 
\usepackage{graphicx}
\usepackage{subcaption}
\usepackage[linesnumbered,lined,ruled,commentsnumbered,algosection]{algorithm2e}
\usepackage{algpseudocode}
\SetKwInput{KwInput}{Input}
\SetKwInput{KwOutput}{Output}

\begin{document}

\title{Heuristic and Reinforcement Learning Algorithms for Dynamic Service Placement on Mobile Edge Cloud}

\author{Dhruv Garg}
\authornote{Based on work done as a Guest Researcher at Ericsson Research, India}
\affiliation{%
\institution{Georgia Institute of Technology}
  \city{Atlanta}
  \country{USA}
}
\email{dgarg39@gatech.edu}

\author{Nanjangud Narendra}
\affiliation{%
  \institution{Ericsson Research}
  \city{Bangalore}
  \country{India}
  }
\email{nanjangud.narendra@ericsson.com}

\author{Selome Tesfatsion}
\affiliation{%
  \institution{Ericsson Research}
  \city{Stockholm}
  \country{Sweden}
}
\email{selome.kostentinos.tesfatsion@ericsson.com}

\renewcommand{\shortauthors}{D. Garg, N. Narendra and S. Tesfatsion}

\begin{abstract}
  Edge computing hosts applications close to the end users and enables low-latency real-time applications. Modern applications in-turn have adopted the microservices architecture which composes applications as loosely coupled smaller components, or services. This complements edge computing infrastructure that are often resource constrained and may not handle monolithic applications. Instead, edge servers can independently deploy application service components, although at the cost of communication overheads. Dynamic system load in mobile network cause like latency, jitter, and packet loss to fluctuate frequently. Consistently meeting application service level objectives while also optimizing application deployment (placement and migration of services) cost and communication overheads in mobile edge cloud environment is non-trivial. In this paper we propose and evaluate three dynamic placement strategies, two heuristic (greedy approximation based on set cover, and integer programming based optimization) and one learning-based algorithm. Their goal is to satisfy the application constraints, minimize infrastructure deployment cost, while ensuring availability of services to all clients and User Equipment (UE) in the network coverage area. The algorithms can be extended to any network topology and microservice based edge computing applications. For the experiments, we use the drone swarm navigation as a representative application for edge computing use cases. Since access to real-world physical testbed for such application is difficult, we demonstrate the efficacy of our algorithms as a simulation. We also contrast these algorithms with respect to  placement quality, utilization of clusters, and level of determinism. Our evaluation not only shows that the learning-based algorithm provides solutions of better quality, it also provides interesting conclusions regarding when the (more traditional) heuristic algorithms are actually better suited.
\end{abstract}




\maketitle

\section{Introduction}


The area of cloud computing is being revolutionized by the shift towards the edge. This is also enabled by the shift of traditional monolithic cloud applications into chains of microservices ( interchangeably referred to as \textit{services} in this paper), which can be independently deployed to run on different edge servers. While edge computing significantly reduces the round-trip time (latency) and jitter (variation in latency), SLO attainment can vary based on dynamic demand and network conditions. The application control logic monitors the system behavior and dynamically scales up or down the services to provide optimum performance. Infrastructure deployment cost however raises the key issue of optimally placing microservice chains.

The issue becomes especially acute for applications deployed over \emph{mobile edge networks}, which have their own challenges. First, edge sites are resource constrained, which means that some sites may not be capable to run certain services. Second, service replication (duplicate service instances) may be needed across edge sites. Depending on the service, this may be costly or even infeasible. However, there is trade-off since service replication might be needed to meet SLOs and excess replication adds to redundancy and monetary cost. Third, system dynamics could necessitate service migration. This includes network changes or faults in any part of the system (e.g. congestion leading to performance degradation) and UE movement between zones controlled by different edge servers (migrating services/ sessions to \textit{follow} users and ensure minimal latency). It is thus a multi-objective problem to meet application SLAs and user experience while minimizing cost. 

In general, dynamic service placement for mobile edge cloud can be of two types: heuristic and learning-based. Heuristic methods wait to receive triggers before taking a service replication, migration or eviction decision. Learning based methods are used when there is a possibility of further optimizing the system based on learning from the history of aforementioned triggers.

In this paper, we present three algorithms for dynamic service placement in the mobile edge cloud- two heuristic (set-cover greedy optimization and integer programming), and one learning-based (reinforcement learning ~\cite{Fan_2020}). We mention the various application requirements and deployment constraints handled by the algorithms, and describe how the algorithms address them. 

For our experiments, we focus on edge computing based low-latency real-time applications. DeathStarBench ~\cite{gan2019open}, is a benchmark covering representative use cases for different application types. They also demonstrate the autonomous drone swarm navigation application which aligns with our edge-computing workload requirement. Thus, we model this particular use case for our experiments to test and compare the three implemented algorithms based on certain quantitative and qualitative metrics. Our results demonstrates that although the learning-based algorithm generates the most optimal placement result compared to the other two, there are situations where the heuristic algorithms do perform quite well too. Further, the results provide insights and reference for practitioners while choosing a placement algorithm for their application.

In particular, our contributions are:
\begin{enumerate}
\item Design and implementation of three multi-objective optimization algorithms to deploy services on mobile edge network, while satisfying constraints. In particular, our placement algorithms are- to the best of our knowledge, \emph{the first such algorithms to take the underlying mobile network infrastructure into account}.

\item Analysis of the algorithm solution based on placement quality, infrastructure deployment cost, determinism of results and algorithm execution time.

\item Comparative evaluation of algorithms vis-a-vis each other.

\end{enumerate}


The rest of this paper is organized as follows. In Section \ref{sec:example}, we discuss the drone swarm navigation use case, which also forms the basis for experiments. The heuristic and learning based algorithms are described in Section~\ref{sec:algos}. Results of the experimental evaluation of the three algorithms, and the comparative analysis, are presented in Section~\ref{sec:expts}. Comparisons and directions from the related work are presented in Section~\ref{sec:related}. Finally, the paper concludes with suggestions for future work in Section~\ref{sec:conc}.
\section{Application use case}\label{sec:example}

\par The microservice architecture used in our edge-computing application use case is based on the autonomous drone swarm coordination scenario from DeathStarBench ~\cite{gan2019open}. This is an open-source benchmark suite built with microservices that is representative of large end-to-end services, and is modular and extensible. DeathStarBench includes a social network, a media service, an e-commerce site, a banking system, and IoT applications for coordination control of UAV swarms. 

Most of the top-of-the-line commercially available drones (DJI Mavic Air 2, Yuneec H520, Kespry Drone 2.0, Autel Evo, Skydio 2) are powered by 2500-5200 milli-ampere hour (mAh) lithium batteries and have flight times between 25-35 minutes. Any additional computational tasks (e.g. video analytics or object detection) placed on the drones will further reduce the flight time. Thus, we assume that existing drones do not have sufficient battery power or on-board hardware for such computationally heavy tasks, and need to be offloaded to edge servers. The drones are only responsible for perceiving the environment, sending the captured data to servers and implementing the navigation instructions from the controller. Applications consuming the drone video feeds for further analytics tasks leverage the computational capabilities available at the edge and cloud servers.

\begin{figure}[]
    \centering
    \includegraphics[width=9cm]{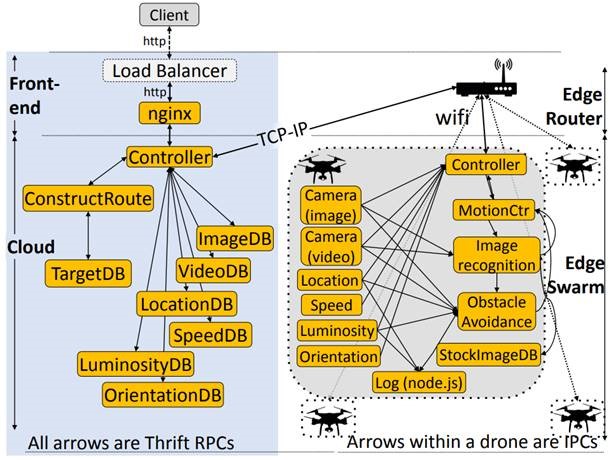}
    \caption{Microservices in drone swarm mobilization using edge resources \cite{gan2019open}}
    \label{fig:droneUML}
\end{figure}

\par The drone swarm mobilization application's microservice chains, taken from~\cite{gan2019open} and their constituent services are depicted in Fig.~\ref{fig:droneUML}. It performs motion planning and coordinated routing for a fleet of drones. It uses location, speed, orientation data, and performs object recognition and obstacle avoidance using the camera feeds. These tasks are compute and storage intensive, and data feeds from the drones are continually processed at the edge servers. The application is also time-critical since it deals with the real-time navigation of currently in flight drones. Given the limited battery power and compute resources on the drone, a viable solution is to transmit and process data at the edge site servers. Due to the drone's proximity with the edge sites, the mobile network can provide a reliable, single hop, high-bandwidth and low-latency communication channel between the edge servers and the drones. While most of the processing is performed at the edge servers, a cloud server can optionally be used for persisting copies of data or constructing the initial route per drone. 

\par Since the application relies heavily on edge servers, its processing throughput will also be limited by their hardware resources. To optimize deployment costs, and meet the strict quality of service (QoS), tasks must be intelligently assigned on servers using the placement algorithm. The complexity of deploying the drone coordination application is further increased by the fact that the navigation services must be accessible to the drones irrespective of their movement across coverage area of individual edge sites. In such scenarios, inefficient placement strategies could lead to over-provisioning of microservices across edge servers and low utilization of server capacity. This is costly and also limits the capacity of edge servers to execute more tasks in parallel. Furthermore, it has been seen that certain placement strategies \cite{svard2015continuous, lopez2018virtual, sedaghat2016decentralized} come up with optimal static placements for a \textit{given system state}, but are be unable to dynamic situations. In case of network, UE mobility or failure events, they make ad-hoc runtime decisions which increase the deployment cost.

\par These challenges of resource management in a edge-centric heterogeneous deployment, strict QoS, requirement of service access across a coverage area, and moving UE (drones) make this application a fit use-case to test our placement algorithms.
\section{Dynamic Placement Algorithms}\label{sec:algos}





\subsection{Overview}\label{subsec:overview}




In this section, we present the three dynamic placement algorithms for edge computing applications based on microservices architecture. These algorithms are generalizable to any network topology and application scenario. Due to time constraints, we test the algorithms for the drone swarm coordination application (described in section ~\ref{sec:example}) alone, but the algorithms have been designed and implemented to remain application agnostic. The first algorithm is a traditional heuristic weighted set-cover approach~\cite{hassin2005better}. The second is also a heuristic algorithm based on a mixed integer linear programming approach as taken by ~\cite{bahreini2017efficient}. Finally, the third algorithm is a reinforcement learning approach developed using the well-known OpenAI Gym environment~\cite{brockman2016openai} and uses the Proximal Policy Optimization (PPO)~\cite{schulman2017proximal} algorithm. 


\subsubsection{Input and output of the algorithms}\label{subsubsec:inputs}
\begin{figure}[]
 \centering
 \includegraphics[width=9cm]{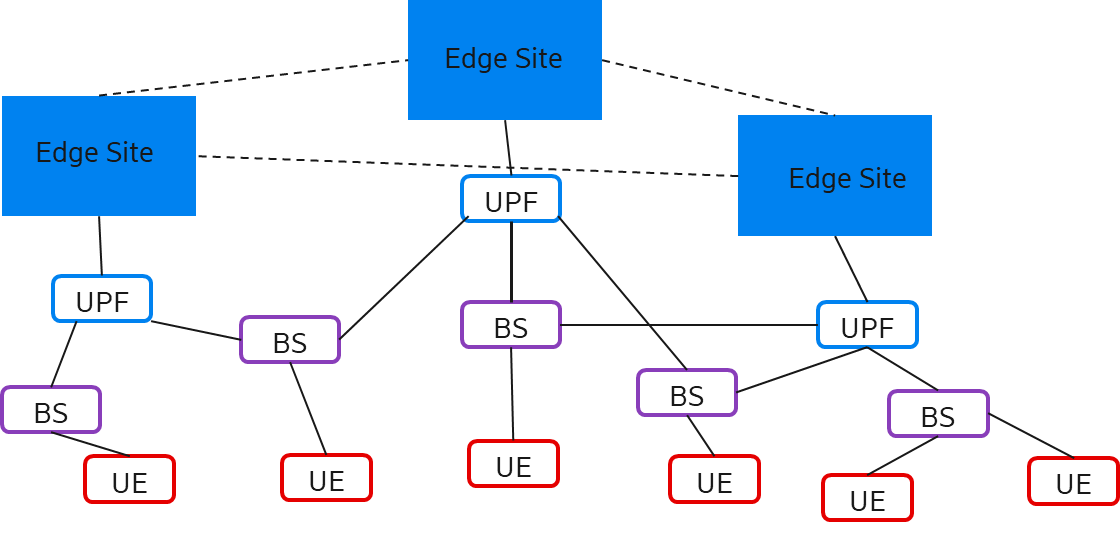}
 \caption{An example network connectivity of mobile network and edge sites}
 \label{fig:mntopology}
\end{figure}

\textbf{Input: } Two sets of inputs are used; service description graph containing information about the service (e.g., microservice chains, set of their constituent services, resource requirements, QoS requirements \textit{communication overhead, data locality constraints, microservice collocation constraints, service requirement constraints }) and a network graph describing the mobile network topology (e.g. connections between the UE and Base Station, links between the BS and the user plane anchor point (e.g., the User Plane Function in 5G, latency links etc. For better understanding, a table of data attributes of the configuration file are given below. \textit{TODO: Add a table of the required inputs, with attributes marked as required or not}. For instance, Figure \ref{fig:mntopology} illustrates three edge sites that are deployed at the edge of a 5G network. Each edge site is connected to an User Plane Function (UPF) via N6 interface. The UPF is the network function providing IP Anchor Point for the UEs in 5G network. One UPF can be connected to multiple BSs via N3 interface. And at a given time, the UE is connected to one BS. However, it is possible that one UE can access multiple edge sites through different UPFs. 
Each edge site has network connections with other edge sites. The network link latency of all links in Figure~\ref{fig:mntopology} can be determined and be made accessible to other entities in the mobile network and the edge infrastructure. In addition to the server sites, base station, and UPFs, the network topology file contain information such as node neighbors, total capacity and unit cost of resource of edge sites, and delays, source, and destination of communication link. The monitored available capacity of edge sites are also fed to the algorithms. 

\textbf{Output: } The output for each algorithm is a set of replicas (or instances) of each microservice and their placement on edge servers in the cluster.

\begin{table}
\begin{center}
\begin{tabular}{|p{2cm}|p{6cm}|}
 \hline
 \textbf{Term} & \textbf{Definition} \\
 \hline
 \textit{User Plane Function (UPF)} & A software component that supports features and capabilities to facilitate user plane operation. Examples include: packet routing and forwarding, interconnection to the Data Network, policy enforcement and data buffering \\ 
 \hline
 \textit{User Equipment (UE)} & Any device used directly by an end-user to communicate \\
 \hline
 \textit{Base Station (BS)} & Transmitter that relays wireless signals - typically between User Equipment and Edge Sites - using radio frequencies  \\
 \hline 
 \textit{Edge site} & An intermediate cluster of one or more servers typically connected to a Base Station, other Edge Sites and one or more cloud servers\\
 \hline
 \textit{N3 interface} & Interface between the radio access network and the UPF \\
 \hline
 \textit{N6 interface} & Interface betweeen UPF and any external service (or internal) networks or service platforms  \\
 \hline
\end{tabular}\\[1ex]
\caption{Technical terms involved in the network connectivity graph}
\label{table:technicalTerms}
\end{center}
\end{table}

\begin{table}
\begin{center}
\begin{tabular}{|p{3cm}|p{5cm}|}
 \hline
 \textbf{Symbol} & \textbf{Definition} \\
 \hline
 \textit{Graph G(N, L)} & Network graph consisting of N network components and L links  \\
 \hline
 \textit{Network components (N)} & Consists of B = {b1-bn} BSs, U = {u1-un} BSs, E = {e1-en} edge sites \\
 \hline
 \textit{Network links (L)} & Consists of L = {l1-ln} network links between nodes in N \\
 \hline
 \textit{Microservices (S)} & Set of microservices in the application, S = {s1-sn}  \\
 \hline
 \textit{Chains (C)} & Larger service chains C={c1-cn} made by the combination of microservices S  \\
 \hline
 \textit{Latency requirements (R)} & Application given SLOs for latency limits R={r1-rn} of each chain in C  \\
 \hline
 \textit{Placement ($\Psi$)} & Mapping between e in E and s in S \\
 \hline
\end{tabular}\\[1ex]
\caption{Definitions for symbols used in the algorithms}
\label{table:algoTerms}
\end{center}
\end{table}

\subsubsection{Objective Function}\label{subsubsec:objfn}
The primary goal of the algorithms is to ensure that availability of all application services at all user equipment (UEs) e.g. in our case drones that can access various application microservices from within the coverage area. In addition to ensuring service access, the algorithms aim to minimize the deployment cost while also meeting the application requirements defined in the scenario. The algorithms, supported costs and constraints are described below.

\subsection{WSSP: Weighted Set-cover based Service Placement}
The rationale behind weighted set-cover is to find a cover with the smallest subsets of a collection of sets over a universe whose union equals the universe and whose total weights are minimum~\cite{hassin2005better}. Our WSSP tries to minimize the weighted sum of costs such as deployment, processing, communication, and storage costs while fulfilling requirements of service performance and capacity constraints of sites in the distributed edge. As output it produces the smallest number of edge sites and their mapping to individual microservices that meets the latency limits for all UEs accessing the service. The algorithm, Algorithm~\ref{alg:hWSSP}, begins by sorting the microservice chains in ascending order by latency requirement (line 1). It then adds edge sites to the list of potential edge sites $P$ for placement if latency requirement of a chain $c_{i}$ can be satisfied from any base station (BS) (helps transmit information between the UE and the edge site) to the edge sites and edge sites have enough capacity to admit all microservices that belong to $c_{i}$ (lines 2-7). If a site cannot host all microservices, the chain $c_{i}$ is split and potential subset of edge sites for the split microservices from $c_{i}$ are selected (line 9). The details of \textbf{Split-and-assign} can be found in Algorithm \ref{alg:split}. Next, costs/weights to the potential edge sites are calculated using a scoring function (line 10). As the optimization problem has multiple objectives, a scalarization method~\cite{marler2004survey} is used to transform the problem into a single objective optimization problem. The scalarized score function is expressed as: 
\begin{equation}
\begin{split}
F_{score} =( a_{1}F_{Dep}) + (a_{2}F_{CPU}) + (a_{3}F_{storage}) \\ 
+ (a_{4}F_{comm}) + (a_{5}F_{update})
\end{split}
\label{eqn:score}
\end{equation}

where $F_{score}$ is a weighted sum of the normalized deployment ($F_{Dep}$), CPU ($F_{CPU}$), storage ($F_{storage}$), communication ($F_{comm}$) and update ($F_{update}$) costs. $a_{1}, a_{2}, a_{3}, a_{4}, and a_{5}$ are the weights for the importance of deployment, CPU, storage, communication and update costs, respectively and $a_{1}+ a_{2}+ a_{3}+ a_{4}+ a_{5}$=1.
The   parameter values need to be chose in order   to achieve the desired trade-offs. Finally, the algorithm finds the minimal number of edge sites for $c_{i}$ that can satisfy the resource and latency constraints for any BS using \textbf{Find\_minimal\_sites} algorithm (refer Algorithm \ref{alg:WSSP}).

\begin{algorithm}
\caption{WSSP Algorithm}
\label{alg:hWSSP}
\SetAlgoLined
\KwInput{Sets of G, S, C, R as defined in table \ref{table:algoTerms}} 

\KwOutput{Placement mapping $\Psi$}
Sort C in ascending order by latency requirements \\
\For{\texttt{$c_{i}$ in C}}{
Let P represent potential edges sites for placement. \\
\For{\texttt{$e_{j}$ in E}} {
  Select $e_{j}$ if $r_{i}$  can be satisfied from any b $\in {B}$ to $e_{j}$ \\ \eIf{$e_{j}$ has enough capacity for $c_{i}$}{$P \leftarrow P \cup e_{j}$ } 
  { $P \leftarrow P \cup $
  \textbf{split-and-assign($c_{i},r_{i},e_{j}$)} \\
  Assign weights W to P (using score function using Eq \ref{eqn:score}) \\
  \textbf{Find-minimal-sites(W,P)}
  }
}
}
\end{algorithm}

To find a split list of microservices of a chain $c_{i}$ and associated selected edge sites, Algorithm~\ref{alg:split} uses a metric as a criteria for splitting (line 1). Depending on the objective, the metric can be based on the demand of individual microservices and available capacity of the site to be selected, dependency among microservices, and collocation constraints or a combination of them. Here we have used a simple approach that splits microservices based on microservice demand and available capacity of edge clusters. For the cluster that is selected to host the first part of the split microservices, a set of neighboring sites are sorted by their latency related to it in ascending order (line 2-4). If latency and resource requirements can be met, the remaining part of $c_{i}$ is assigned to a site that is being evaluated (line 6-8). If resource requirement cannot be met then the same algorithm is called again with the remaining part of $c_{i}$ and the neighbour site as inputs (line 10). 


\begin{algorithm}
\SetAlgoLined
\KwInput{Chain $c_{i}$,  latency requirement of  $c_{i}$ $r_{i}$, edge site $e_{j}$,
}

\KwOutput{Split list of microservices $s_{i}$, selected edge sites $e_{i}$}
Split $c_{i}$ based on metric \\
Assign the first part of $c_{i}$ to $e_{j}$ \\
$C_{r}\leftarrow {c_{i}\_remaining\_part}$ \\
$neigboringSet_{e_{j}}$ $\leftarrow$ Find a set of $e_{j}$’s neighbor sites, and sort by latency in ascending order \\
\For{\texttt{$k \in neigboringSet_{e_{j}}$}}{
 \eIf{$r_{i}$ can be met}{\eIf{resource requirement can be met}{assign $C_{r}$ to k}{split-and-assign($C_{r}$,k)}}{Continue}
  
}
\caption{Split-and-assign($c_{i},r_{i},e_{j}$)}
 \label{alg:split}
\end{algorithm}

Algorithm~\ref{alg:WSSP} is based on the weighted set-cover algorithm ~\cite{hassin2005better}. It finds the minimum number of sites for microservice placement to provide microservices within the required latency limit for all UEs. The basic idea is to select sites with minimum ratio of weights ($W_{i}$) and number of added sites ($S_{i}$) (see line 4-5).

\begin{algorithm}
\SetAlgoLined
\KwInput{A collection of subset of the universe BS covered by potential edge sites, S = $\{S_{1},S_{2},...S_{m}\}$, Weights of elements of S, W = $\{W_{1}, W_{2}, ... W_{m}\}$
}
\KwOutput{Minimum number of cost effective edges sites that cover all elements of BS }
Let I represent set of elements included so far. \\ 
Initialize I = \{\} \\
\While{\texttt{I is the same as BS}}{ 
Choose $S_{i} \in {S}$ minimizing the ratio of the weight $W_{i}$ and number of newly added elements, i.e., Wi / $|Si - I|$. \\ 
Add elements of above picked $S_{i}$ to I, i.e., $I$=$I$ $\cup$ $S_{i}$}
Return I
 \caption{Find-minimal-sites(W,P)}
 \label{alg:WSSP}
\end{algorithm}

\subsection{MISP: Mixed-Integer linear programming based Service Placement}
The MISP placement algorithm is executed in three stages. Initially, the algorithm evaluates cost of deployting microservices to servers based on compute resources and constraints. The resulting matrix is solved using the mixed integer linear programming solver to return an intial placement mapping each microservice to one server. Next, microservice communication costs are taken into account. MISP tries to minimize these costs by deploying communication-heavy microservices closer to their dependent microservices. These updates can reduce the overall placement cost. In the third stage, MISP checks for SLO attainment and if all UEs are able to access the services. In case some UEs report one/more services exceeding the SLO, MISP deploys more microservice instances closee to these UEs.

Our algorithm improves upon the MCAPP-IM ~\cite{bahreini2017efficient} algorithm in four ways. First, based on microbenchmarks, we found that Hungarian-matching took three to four orders of magnitude greater time to solve larger systems of equations i.e. larger network and service chains. Hence, we replaced the hungarian algorithm as the solver with MISP. Second, the MCAPP-IM algorithm assumed that only one microservice would be assigned to any server. This is not realistic and server utilization needs to be maximized. Third, MISP supports a number of real-world application constraints which include service constraint (microservices placed on certain servers due to hardware requirements), collocation constraint (tightly coupled microservices that need to be placed on the same server) and data locality constraints (sensitive user data that must be stored on servers within a defined region). The MISP algorithm is described in Algorithm~\ref{alg:misp}.

\begin{algorithm}[] 
\SetAlgoLined
\KwInput{Sets of G, S, C, R as defined in table \ref{table:algoTerms}, For $t$ > 0, solution $\Psi_{t-1}$ is also given}
\KwOutput{Placement solution $\Psi_{t}$}
Read application file to extract computation ($\gamma$), constraints and communication costs ($\delta$)\\
Populate $\tau$, containing cost for deploying the microservices across edge sites \\
Obtain $\Psi_{mip}$ using \textbf{mipSolver} which places all microservices on exactly one server \\
Calculate the initial cost $\Psi_{mip,cost}$ using \textbf{calcCost} \\
New placement $\Psi_{ls}$ using \textbf{LSearch} to reduce communication overhead \\
Get final $\Psi_{t}$ after calculating \textbf{serviceLatency} for each UE
\caption{MISP Algorithm}
\label{alg:misp}
\end{algorithm}

\par The initial microservice-server placement is done based on deployment cost alone (lines 1-3, Algorithm \ref{alg:misp}). For this, application and topology data is read and costs for $\gamma$ (compute), $\delta$ (microservice-user communication), and $\rho$ (microservice migration) are computed. These values help formulate the cost matrix, i.e. TMatrix ($\tau$) in line 2. $\tau$ is a $k x n$ matrix where $k$ is the number of microservices and $n$ is the number of servers. 
Each row in $\tau$ represents a microservice $s$, and columns represent its deployment cost on different servers $e$. This factors in heterogeneous edge compute where the deployment costs can vary based on the server hardware. 
$\tau$ adds or ignores $\rho$ (microservice relocation) costs based on whether a microservice can be migrated from server $e_{i}$ to $e_{j}$. The TMatrix also enforces the application's service ($svc_{c}$) and data locality ($loc_{c}$) constraints. It does so by determining microservice-server placements which are not allowed by the application and assigns those costs to a large numeric constant, $\kappa$. The solver (which takes $\tau$ as input) naturally eliminates solutions with such server-microservice combinations as they would significantly increase the placement cost.


\par The TMatrix ($\tau$) and edge site capacity ($e_{cap}$) constraints are given as input to \verb|mipSolver| which uses Google's assignment solver\footnote{https://developers.google.com/optimization/assignment/overview}. Based on these inputs, we construct constraints for the MIP solver to generate the initial placement solution, mapping one microservice to exactly one server while minimizing the overall cost. This gives us the initial microservice placement solution ($\Psi_{mip}$), based on the compute and relocation costs, application constraints and the edge server resource limits. 


Second, a heuristic search \verb|LSearch| (Algorithm~\ref{alg:LSearch}) is used to reduce the inter-microservice communication costs in $\Psi_{mip}$ solution, obtained in (line 3, Algorithm \ref{alg:misp}). It computes $\Lambda$, the communication costs incurred by each of the microservices based on their interaction with other microservices ($s_{ij}$) and current server placement in $\Psi_{mip}$. LSearch evaluates combinations to move heavily communicating microservices closer to each other since this could reduce the communication cost and the overall deployment cost. However, we note that this migration of microservices is only allowed where it is permitted based on defined application constraints. Costs for the different combinations ($\Psi_{ls,cost}$) is compared with $\Psi_{mip,cost}$. If $\Psi_{ls,cost}$ is lower, the placement solution is updated, else, the previous solution is retained.


\begin{algorithm}[] 
\SetAlgoLined
\KwInput{Initial placement $\Psi_{mip}$ and its cost from \textbf{mipSolver}}
\KwOutput{$\Psi_{ls}$: Placement after LSearch}
Compute communication cost using $\Psi_{mip}$ placement and microservice dependencies \\
Find microservices with highest communication overheads \\
Try more placement combinations to reduce communication cost of identified microservices \\
Update placement solution if new placement combination reduces overall cost \\
\caption{LSearch()}
\label{alg:LSearch}
\end{algorithm}

\par Third, MISP ensures availability of application services are provided within latency limits, to all the UEs in the coverage area. \verb|serviceLatency| (Algorithm ~\ref{alg:serviceLatency}) invokes \verb|calcLatency| to calculate service chain latency for $all$ the services provided at $each$ UE. As mentioned previously, service latency for a UE is the sum of all network links in the graph, connecting the UE to the nearest servers hosting the microservices required for that service. Duplication of microservices
on servers close to the UE is triggered if a service chain is not provided within the latency limit (line 14, Algorithm~\ref{alg:serviceLatency}). To provide maximum services with minimum deployment cost, the microservice duplication is done in ascending order of compute load (line 13, Algorithm~\ref{alg:serviceLatency}), and at the server connected to the UE with lowest latency link. To reiterate, the new deployment of a microservice $s$ on server $e$ is dependent on the application constraints. After each microservice duplication, service latency to UEs is re-computed. The duplication step is repeated until SLO is met or there are more combinations to try. At this stage, we have the final placement solution $\Psi_{t}$.

\begin{algorithm}[] 
\SetAlgoLined
\KwInput{$C$, $E$, $B$, $\tau$, $\Psi_{ls}$, $G$, $colloc_{c}$}
\KwOutput{$\Psi_{t}$: Final placement after latency check and duplication}
Iterate over base stations B \\
For each edge server calculate the $shortest\_path(G,b,e)$ network latency to the base station \\
Iterate over service chains for each base station B \\
Use \textbf{calcLatency} to determine total network latency to provide chain $c$ to BS $b$ \\
If current latency $r$ lesser than permissible limit, continue \\
Else, use \textbf{duplicate} to greedily create new instances of a microservice $s$ on a nearby server \\
Update $\Psi_{t}$ with newer $\Psi_{t'}$ after every placement change
\caption{serviceLatency()}
\label{alg:serviceLatency}
\end{algorithm}

\subsection{RLSP: Reinforcement learning based Service Placement}



\par Reinforcement Learning (RL)~\cite{mitchell1997machine} is a machine learning technique where agents learn a \textit{policy} from trial and error, in an interactive environment. The concepts of \textit{history} and \textit{state} are central to it. The \textit{history} describes all the interactions that have taken place between the environment and the agent, and its abstract representation is called \textit{state}. Key concepts of RL include delayed rewards (short term and long term rewards), importance of time (state at time $t+1$ is dependent on the state at time $t$) and that an agent's action affects its next input, next action and its future path. The RL agent learns about the environment by repeatedly taking actions and modifying states based on the reward obtained. The goal of the RL agent is to maximize the overall reward. We designed and implemented a custom OpenAI Gym \cite{brockman2016openai} environment, called RLSP (Algorithm ~\ref{alg:rl_env}), to model the service placement problem as a Reinforcement Learning (RL) problem.


\begin{algorithm}[] 
\SetAlgoLined
\KwInput{Observation $obs$ representing the current environment state and the $reward$ value}
\KwOutput{An $action$ to perform on the environment to maximize $reward$ while satisfying service latency, application and resource constraints}
\textbf{\_init\_(self)} \\
\hskip0.5em Define environment variables, action\_space and observation\_space \\

\textbf{\_next\_observation(self)} \\
\hskip0.5em Loads the next frame and provides environment information to the RL agent \\

\textbf{step(self, action)} \\
\hskip0.5em Performs an action and obtains reward and observable information from the environment \\

\textbf{\_take\_action(self, action)} \\
\hskip0.5em Performs an action on the environment \\

\textbf{reset(self)} \\
\hskip0.5em Reset values in all variables and matrices \\

\textbf{render(self)} \\
\hskip0.5em Display current environment state, reward and counter
\caption{RLSP: An RL agent for Service Placement}
\label{alg:rl_env}
\end{algorithm}

\par The algorithm includes an RL agent which has (i) \textit{action\_space} and \textit{actions}, (ii) \textit{observation\_space}, and a \textit{reward function}. In the \verb|init| method (Algorithm ~\ref{alg:rl_init}), the application and topology data are read, and the cost matrix i.e. TMatrix ($\tau$) is computed. Here, we also define the shape and type of the environment's \textit{action\_space} and \textit{observation\_space}. An action by RLSP includes \textit{three} attributes: (i) action type (deploy, evict or hold a microservice), (ii) microservice index [$0,s_{j}-1$] where $s_{j}$ is the number of microservices in the application, and (iii) server on which the action has to be taken [$0,e_{n}-1$], where $e_{n}$ is the number of servers. Thus, the action space (actions taken by the agent) is a list of three values where each value is a discrete integer within the defined range. The observation space (environment information) gives a structure for the environment to return information on (i) the number of service chains [$0,C_{k}$] ($C_{k}$ is the number of service chains in the application) accessible by each of the $b_{l}$ BS (base stations), and (ii) the number of microservices [$0,s_{j}$] deployed on each of the $e$ servers. Thus the observation space is a list of size [$B+E$]. Since both \textit{action\_space} and \textit{observation\_space} take discrete integer values in a defined range, they are of \verb|MultiDiscrete| type. Besides this, init also initializes variables for \verb|reward|, \verb|counter|, \verb|serverMicroservices| (matrix of microservices deployed on servers) and \verb|userServicesAccess| (matrix of service chains accessible from base stations).

\begin{algorithm}[] 
\SetAlgoLined
\KwInput{$\Psi$: Given placement solution}
\KwOutput{$\Psi_{cost}$: Placement cost}
Initialize $reward$, $rewardOrPenalty$, $counter$, $max\_steps$ and create matrices $servMicro$ and $userServAccess$ \\
Evaluate compute ($\gamma$) and constraint costs from application, topology files \\
Calculate matrix $\tau$ using \textbf{calcT} \\
Define $action\_space$ and $observation\_space$ \\
\caption{\_init\_()}
\label{alg:rl_init}
\end{algorithm}

\begin{algorithm}[] 
\SetAlgoLined
\KwInput{$self$}
\KwOutput{$obs$}
Load the next input frame of application, network data \\
$userServAccess$ $\leftarrow$ \textbf{serviceLatency} ($C$, $E$, $B$, $\Psi$) \\
Populate $obs$ of $observation\_space$ type using $userServAccess$ and $servMicro$
\caption{\_next\_observation()}
\label{alg:rl_nextObs}
\end{algorithm}

\begin{algorithm}[] 
\SetAlgoLined
\KwInput{$self$, $action$}
\KwOutput{$status$, $rewardOrPenalty$} $act\_type$, $act\_micro$, $act\_serv$ $\leftarrow$ $action$ \\
Initialize $rewardOrPenalty$ $\leftarrow$ $0$ \\
\eIf{$action$ is valid based on $servMicro$}{$rewardOrPenalty$ $\leftarrow$ small positive reward \\ 
Update $servMicro$}{$rewardOrPenalty$ $\leftarrow$ small/large negative penalty}
\caption{\_take\_action()}
\label{alg:rl_takeAction}
\end{algorithm}

\begin{algorithm}[] 
\SetAlgoLined
\KwInput{$self$, $action$}
\KwOutput{$obs$, $reward$, $done$}
$counter$ $\leftarrow$ $counter + 1$ \\
$status$, $rewardOrPenalty$ $\leftarrow$
\textbf{\_take\_action}(self, action) \\
$obs$ $\leftarrow$ \textbf{\_next\_observation}() \\
Get $servAvailable$ from $obs$ \\
$cost$ $\leftarrow$ Computed from \textbf{calcCost}($\Psi$) and $\delta$ \\
Get new $reward$ based on $servAvailable$, $cost$, $max\_steps$ and $delay\_modifier$ \\
\eIf{$counter$ $>=$ $max\_steps$}{$done$ $\leftarrow$ $true$}{$done$ $\leftarrow$ $false$}
\caption{step()}
\label{alg:rl_step}
\end{algorithm}

\par The \verb|step| function (Algorithm~\ref{alg:rl_step}) is triggered at each time-step. First, it invokes the \verb|take_action| method to perform an action on the environment. Next, it calls the \verb|next_observation| method to obtain the next state from the environment. 

\par The \verb|take_action| (Algorithm~\ref{alg:rl_takeAction}) parses the aforementioned three values from the input action and also takes the current deployment state from \verb|serverMicroservices|. Any action to be performed on the environment may be valid or invalid, depending on the current state. Valid actions include: hold or evict instruction for a microservice that was running on a server or deploying a microservice which is not currently running on the server. If the combination of the action\_type, microservice and server form a valid action for the current environment state, a small positive reward is given to the RLSP, and the \verb|serverMicroservices| matrix is updated. On the other hand, the RL agent could also take invalid actions. These include: trying to hold or evict a microservice which is not currently running on a server, or trying to deploy a microservice on a server where it is already running. In such cases, a small negative reward is given to the agent and the \verb|serverMicroservices| matrix is left unchanged. This is also the stage where application constraints are enforced by checking the cost matrix $\tau$. Here, the actions taken by the agent, in violation of the application constraints (data locality, service or collocation) are penalized. Upon receiving a penalty, it is expected that RLSP would learn and avoid such actions in the future.

\par After performing the action, \verb|step| calls \verb|next_observation| (Algorithm~\ref{alg:rl_nextObs}). Here, the next input state of the environment is provided to the agent. This function also gets the updated environment state from \verb|serverMicroservices|, which is given as input to the \verb|checkUserLatency| function to calculate the service chain latency at various UEs. After performing the latency check, \verb|serviceUserAccess| matrix is returned, which contains information on which service chains are accessible from which base stations. The latency check returns the \verb|serviceUserAccess| matrix, which contains information on the service chains accessible from the various base stations.

\par The \verb|step| function uses the two matrices to find the step's net reward. Using the \verb|serviceUserAccess| matrix, a total positive reward is computed based on the total service chains accessible from the different UEs. As the number of service chains provisioned increase, the positive component of the reward will also increase. Another factor included in the reward function is the deployment (compute and communication) cost. It is computed using the \verb|serverMicroservices| matrix and it is subtracted from the reward. This is because the placement strategy aims to provide maximum services through minimum microservice instances, i.e., lower deployment cost. For effective learning of the RLSP, this net reward is further scaled using a discount factor. The discount factor diminishes the value of the reward in the initial time-steps, and keeps increasing its weight as the number of time-steps approach the \textit{max\_steps} value. This ensures that the RL agent is not significantly influenced by large short term positive or negative rewards. Instead, it works towards a large cumulative reward in the long term. At this stage, the \verb|step| function returns the next observation and current reward to RLSP.

\par Thus, in order to maximize the cumulative reward, the RLSP will try to gain more positive reward by ensuring greater accessible service chains and taking valid actions. It will also try to minimize the negative reward from deployment costs and invalid actions. The implementation also has helper functions of \verb|reset| and \verb|render| (Algorithm~\ref{alg:rl_env}) to re-initialize all variables after the current iteration ends, and to display the current system state.

\par Finally, we train and test the RLSP. To train our RL agent, we use the \verb|Stable Baselines3| ~\cite{stable-baselines3} library. This library provides reliable PyTorch implementations of state-of-the-art reinforcement learning algorithms, and a framework to train, test and save RL agents built on the OpenGym environment. In particular, we use the Proximal Policy Optimization (PPO) ~\cite{schulman2017proximal} algorithm to train our agent. The PPO algorithm combines benefits of multiple workers from A2C (Actor-Critic Algorithm), and of trust region from TRPO (Trust Region Policy Optimization). The PPO algorithm uses a Trust Region based objective function which is also compatible with Stochastic Gradient Descent. In particular, PPO provides an ease of tuning and it computes an update at each time-step which not only minimizes the cost function but also avoids drastic deviations from the previous policy. In our experimentation with RLSP, we found that PPO outperformed many other learning algorithms. 
\section{Evaluation}\label{sec:expts}

\subsection{Experimental setup}
\subsubsection{Network topology}
\par We assume a hypothetical telecommunication network setup spanning four geographically distributed edge servers ($E1$ - $E4$), also called edge sites, which are identified by a unique \verb|id| and \verb|lat,long| location. The topology used for the simulation experiments was similar to Figure ref{fig:mntopology}. Their information specified is assumed to include the memory, disk storage, compute capacities, and also the presence of specialized hardware such as Graphics Processing Units (GPU), if any. These resource capacities set an upper bound for the algorithm to only deploy microservices on a server as long as resources are available. The presence of specialized hardware on a server also helps us meet the constraints for microservices that utilize it. For the convenience of our algorithms, we also define a unit cost of storage and compute on a server based on the resources it provides. This plays a role in heterogeneous compute environments where different edge servers provide varying compute power, resulting in different deployment costs. (It is to be noted that the compute/storage capacities of the various edge sites are assumed to be heterogeneous.)
\subsubsection{Application and workloads}
\par The experiments to test the dynamic service placement algorithms use the drone swarm mobilization application, introduced in Section~\ref{sec:example}. This application has 13 service chains which utilize a total of 23 microservices. The service chains were modeled based on the system design given in Figure~\ref{fig:droneUML} and are described in Table~\ref{table:serviceChains}.
\par Each service chain in the application has a configurable latency limit $r_{k}$. The QoS (primarily, latency) requirements of the 13 application service chains are broadly categorized into \textit{ultra-low} or strict (0.2 ms) latency, \textit{moderate} or medium (0.4 ms) latency and \textit{relaxed} or high (0.6 ms) latency. These are typical values for so-called Ultra Reliable Low Latency Communication (URLLC), especially for a critical use case such as drone swarms~\cite{bor20195g}. These 13 chains, in different ratios, are used to create the $Ultra-Low (W1)$, $Moderate (W2)$ and $Relaxed (W3)$ application workloads. Such workloads enable us to assess the placement quality of algorithms when applied to diverse applications with varying QoS requirements. The details in Table~\ref{table:workload} show the count and proportion of \textit{ultra-low}, \textit{medium} and \textit{relaxed} service chains in workloads $W1$, $W2$, and $W3$, respectively. The number of \textit{ultra-low} latency service chains increase their percentage from 30\% to 55\% and further to 70\% as we move from $W1$ to $W2$, and finally to $W3$. These workloads are also modeled to have 20\% microservices with service requirement constraints (e.g. specialized hardware); 10\% microservices with collocation constraint (e.g. tightly coupled microservices which run on the same server) and 20\% microservices with data locality constraint (e.g. sensitive data to be placed only on edge servers in a region).

\subsubsection{Hardware used}
\par We use simulation to conduct experiments on an Intel Core i5-8350U $@$ 1.70 GHz machine with 16 GB RAM. One of the reasons to do simulations is because of the limited access to actual compute and networking testbeds to deploy microservices based application, capture network metrics and apply our 3 algorithms. Secondly, there are very few reinforcement learning frameworks for both training and testing RL agents. If at all a physical environment is created to train an agent for a policy (e.g. robotics), they are very application specific. This is also because the actuation metrics for the RL agents are different. Owing to these limitations, and to have a level comparison between the two heuristic and reinforcement learning approaches, we use simulation. 

\begin{table}
\begin{center}
\begin{tabular}{|p{2cm}|p{6cm}|}
 \hline
 \textbf{Service chain} & \textbf{Constituent microservices} \\
 \hline
 \textit{Frontend} & loadBal, nginx, cloud control (cCtrl)  \\
 \hline
 \textit{Controller Cloud} & nginx, cCtrl, consRoute, edge control (eCtrl)  \\
 \hline
 \textit{Controller Edge} & cCtrl, eCtrl, mCtrl, camVid, camImg, loc, speed, lum, orient  \\
 \hline
 \textit{Construct Route} & cCtrl, consRoute, eCtrl, targetDB  \\
 \hline
 \textit{Image} & cCtrl, eCtrl, camImg, imageDB  \\
 \hline
 \textit{Video} & cCtrl, eCtrl, camVid, videoDB  \\
 \hline
 \textit{Location} & cCtrl, eCtrl, loc, locationDB  \\
 \hline
 \textit{Speed} & cCtrl, eCtrl, speed, speedDB  \\
 \hline
 \textit{Luminosity} & cCtrl, eCtrl, lum, luminosityDB  \\
 \hline
 \textit{Orientation} & cCtrl, eCtrl, orient, orientationDB  \\
 \hline
 \textit{Motion Control} & cCtrl, eCtrl, mCtrl, imgRecog, obsAvoid  \\
 \hline
 \textit{Image Recog} & eCtrl, mCtrl, imgRecog, stockImageDB  \\
 \hline
 \textit{Obs Avoidance} & eCtrl, mCtrl, obsAvoid, log  \\
 \hline
\end{tabular}\\[1ex]
\caption{Service chains: Edge-centric drone mobilization}
\label{table:serviceChains}
\end{center}
\end{table}

\begin{table}
\begin{center}
\begin{tabular}{|p{1.5cm}|p{1.5cm}|p{1.5cm}|p{1.5cm}|}
 \hline
 \textbf{Workload} & \textbf{Relaxed chains} & \textbf{Moderate chains} & \textbf{Ultra-low chains} \\
 \hline
 $W1$ & 2 (15\%) & 7 (55\%) & 4 (30\%)  \\
 \hline
 $W2$ & 2 (15\%) & 4 (30\%) & 7 (55\%)  \\
 \hline
 $W3$ & 2 (15\%) & 2 (15\%) & 9 (70\%)  \\
 \hline
\end{tabular}\\[1ex]
\caption{Workloads and their service chain compositions}
\label{table:workload}
\end{center}
\end{table}
\textbf{Comparative evaluation}: The three algorithms WSSP, MISP and RLSP were implemented to provide solutions to workloads $W1$, $W2$ and $W3$. Their solutions are evaluated and compared on four key parameters: placement quality, application deployment, consistency in results and algorithm execution time.

\subsection{Placement Quality}
A placement algorithm's solution quality can be measured by the extent to which it satisfies Quality of Service (QoS) requirements, application constraints and access to microservices across the coverage area. 


\subsubsection{Costs and constraints}
The three algorithms successfully produced solutions that satisfied the QoS requirements in workloads $W1$, $W2$, $W3$. The placement fulfilled the service, collocation and data locality constraints defined for microservices in the drone application. The algorithms also employed mechanisms to reduce the overall placement costs. The only exception here is that WSSP did not take into account collocation and data locality constraints, which it does not support. This results in WSSP deploying some microservices on edge server E1, which are constrained to run only on edge server E3, as shown by MISP and RLSP in Figures~\ref{fig:mscvDeployed1}, \ref{fig:mscvDeployed2} and \ref{fig:mscvDeployed3}.


\subsubsection{Services provided in the coverage area}
The placement solutions from the three algorithms provided 100\% service coverage, i.e., microservices deployed at assigned edge servers were able to provision all services within their latency limits, for all UEs in the network coverage area. MISP performs duplication of microservices on edge servers close to the UE till the QoS latency is met. This check is performed at all UEs, so that 100\% service coverage is ensured. WSSP does not directly perform duplication of microservices until the latency is met. It has knowledge of all the UEs where latency limit can be satisfied, and it then selects and deploys services on the smallest number of edge servers which provide full coverage. RLSP on the other hand gets a positive reward for each service chain that it provides to any UE, within the latency limits. Hence it takes into account the costs/constraints and deploys microservices on edge servers to provide 100\% service coverage, thereby maximizing its reward. 


\begin{figure*}
    %
    \begin{subfigure}[b]{0.49\textwidth}
    \includegraphics[width=\textwidth]{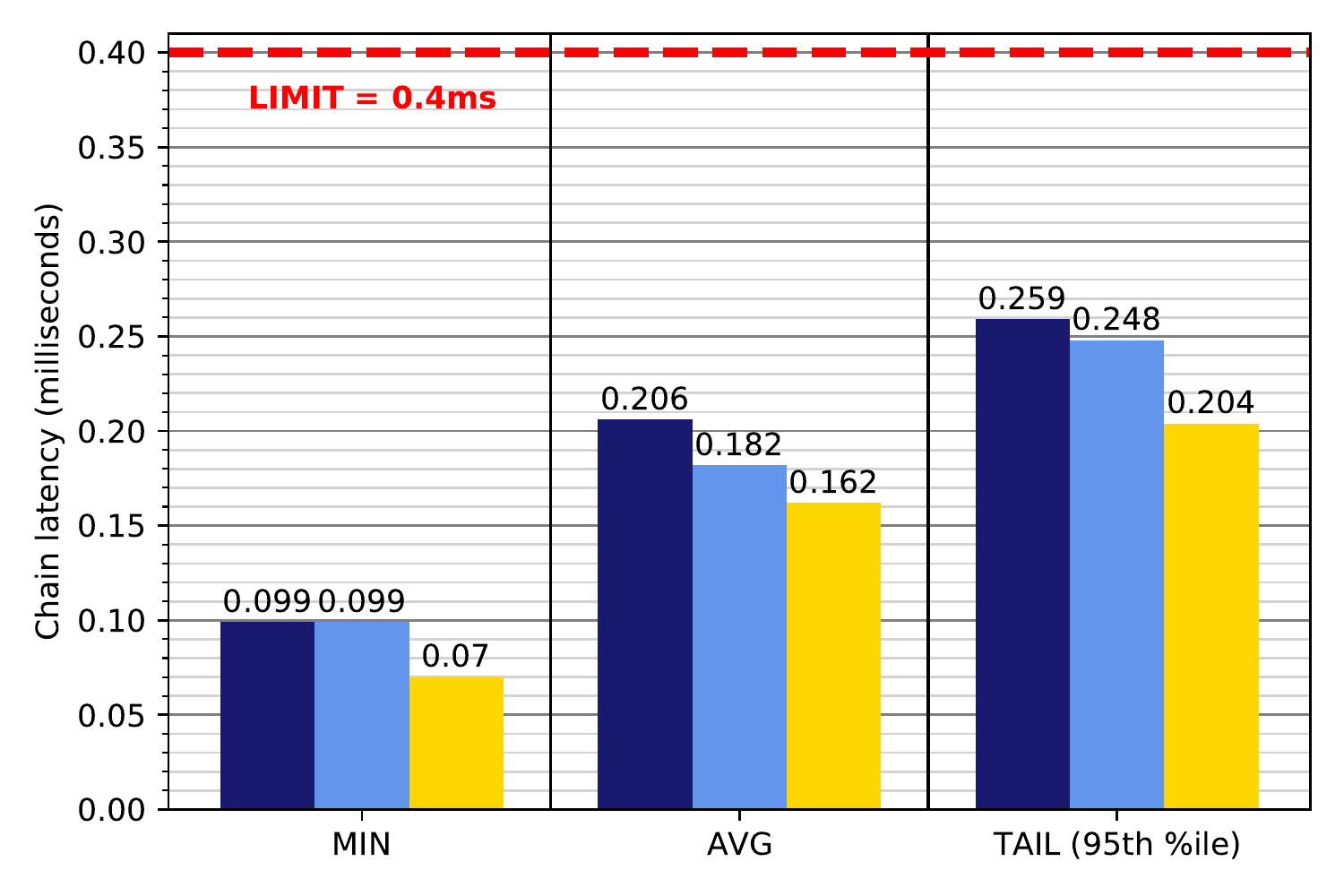}
    \caption{Moderate QoS chain}
    \label{fig:latencyModerate}
    \end{subfigure} \hfill
    \begin{subfigure}[b]{0.49\textwidth}
    \includegraphics[width=\textwidth]{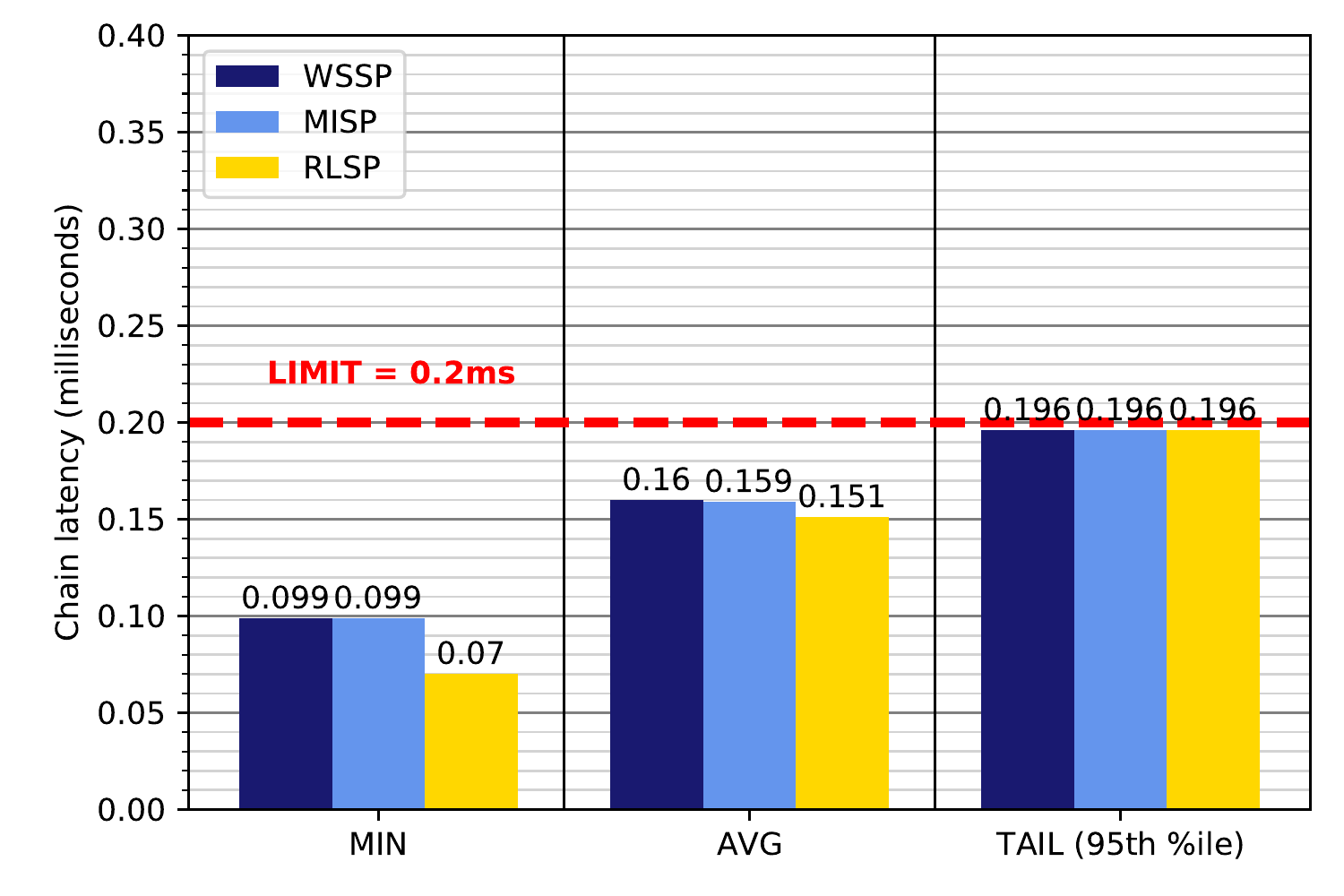}
    \caption{Ultra-low QoS chain}
    \label{fig:latencyLow}
    \end{subfigure}
    
    \caption{Service latency observed at base stations across the coverage area}
    \label{fig:latencyDistribution}
\end{figure*}

\subsubsection{QoS for UE}


\par The 100\% service coverage for the all solutions was verified during experimentation. The distributions for latencies at UEs for two representative moderate and ultra-low service chains are shown in Figure~\ref{fig:latencyDistribution}. These two categories of service chains constitute up to 85\% of the three workloads. In the moderate QoS chain, we see that sufficient deployment of microservice and the latency links between UE and the edge servers ensures that QoS requirements are 35\% to 82.5\% within the 0.4$ms$ latency limit (Figure~\ref{fig:latencyModerate}). For the ultra-low latency service chain, this ranges between the min/avg/tail latencies and 0.2$ms$ limit narrows to 2\% to 65\% (Figure~\ref{fig:latencyLow}), as the MISP and WSSP algorithms make new deployments to ensure that QoS requirements are met. Meanwhile, RLSP manages a lower latency compared to MISP and WSSP due to substantial microservice deployment on $E_{4}$, which is close to many UEs, thereby reducing the end-to-end latency at the UEs.

\subsubsection{Variability of the deployment with evolving conditions}
\begin{figure*}[t]
    \centering
    
    \begin{subfigure}[b]{0.32\textwidth}
    \includegraphics[width=\textwidth]{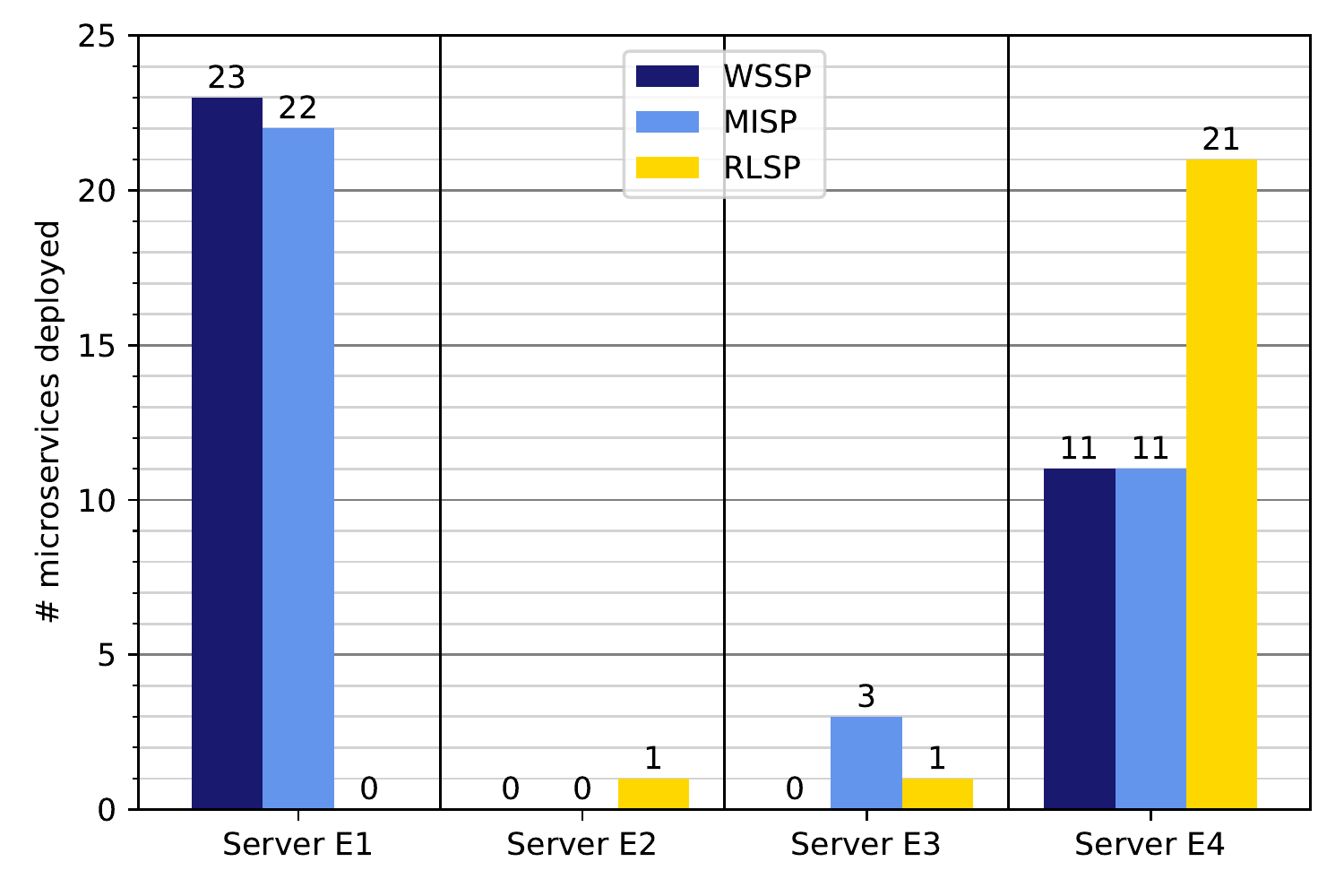}
    \caption{Relaxed workload}
    \label{fig:mscvDeployed1}
    \end{subfigure} \hfill
    \begin{subfigure}[b]{0.32\textwidth}
    \includegraphics[width=\textwidth]{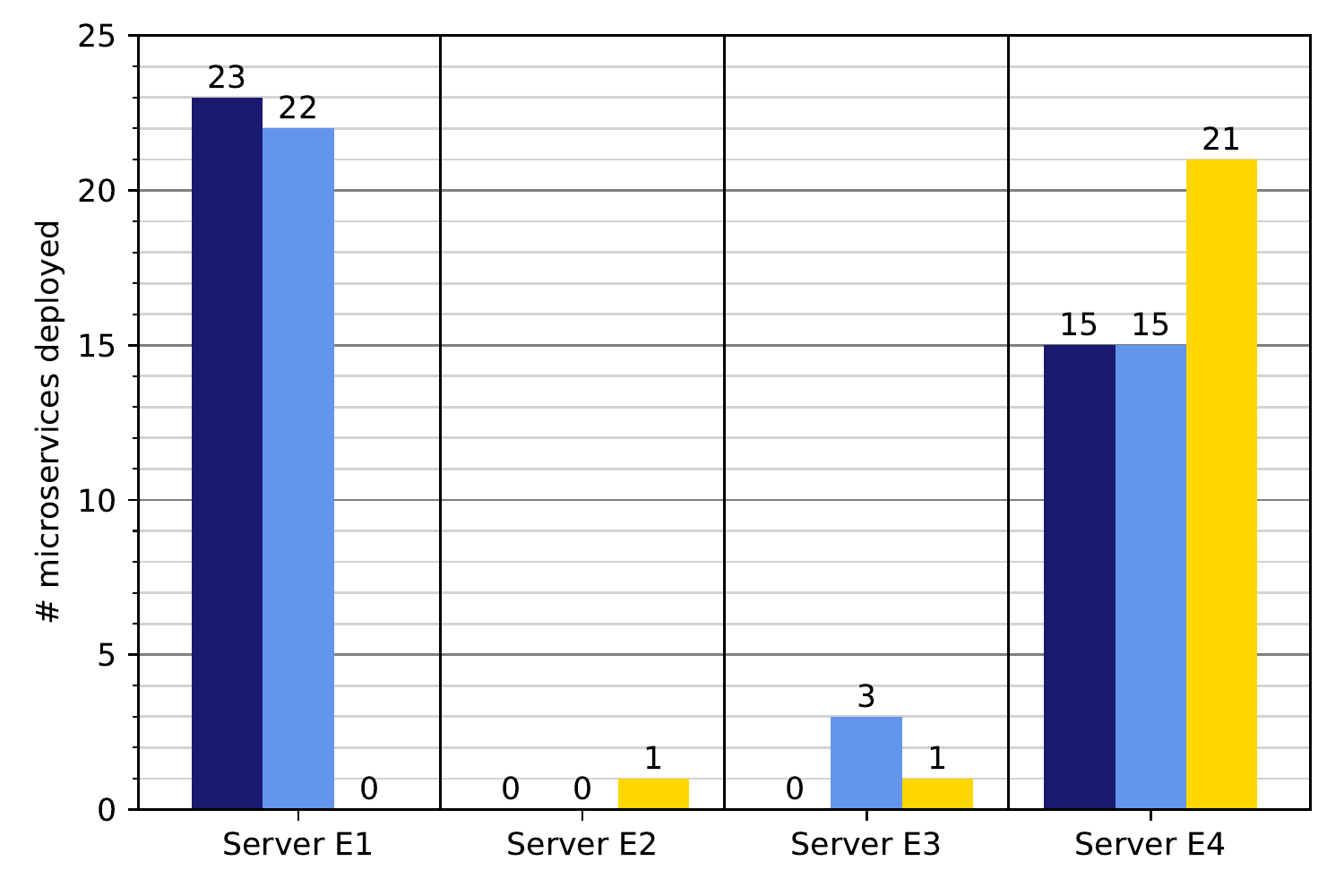}
    \caption{Moderate workload}
    \label{fig:mscvDeployed2}
    \end{subfigure} \hfill
    \begin{subfigure}[b]{0.32\textwidth}
    \includegraphics[width=\textwidth]{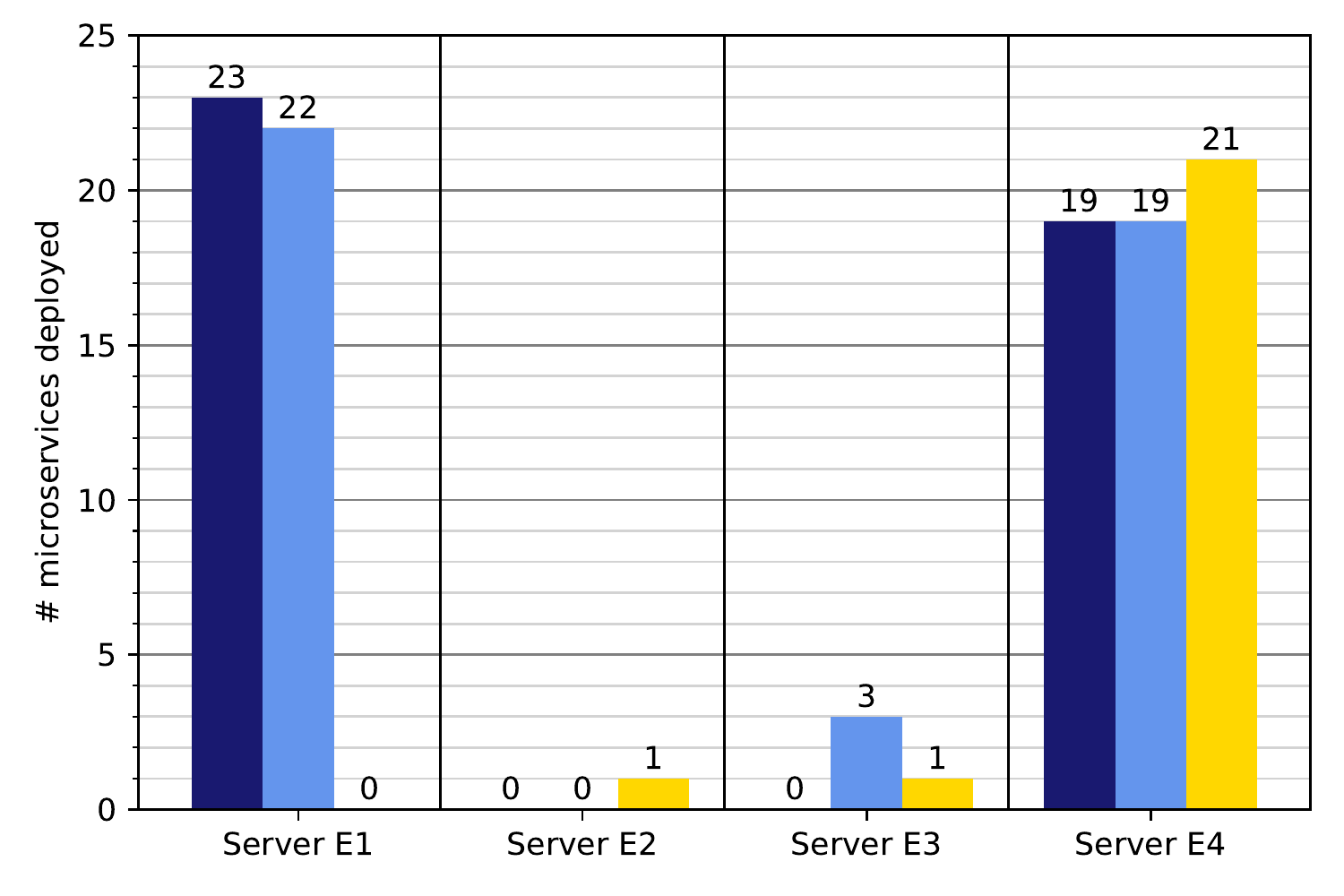}
    \caption{Ultra-low workload}
    \label{fig:mscvDeployed3}
    \end{subfigure}
    
    \caption{Microservices deployed on edge servers across $W1$, $W2$, $W3$}
    \label{fig:microservicesDeployed}
\end{figure*}

Several placement decisions are made by the algorithms to adapt to the changing application requirements and dynamic network conditions. This requires a finite non-zero time and could cause QoS violations while the placement algorithm makes decisions on deploying, migrating or evicting microservices. Figure~\ref{fig:microservicesDeployed} shows the microservices deployed on edge servers by the algorithms for the three different workloads. It can be seen that WSSP and MISP follow a similar trend of retaining the existing microservices on $E1$ while deploying new microservice instances on $E4$ due to the stricter QoS requirements as we move from $W1$ to $W3$. 

\par
\begin{figure}[t]
    \centering
    \includegraphics[width=0.49\textwidth]{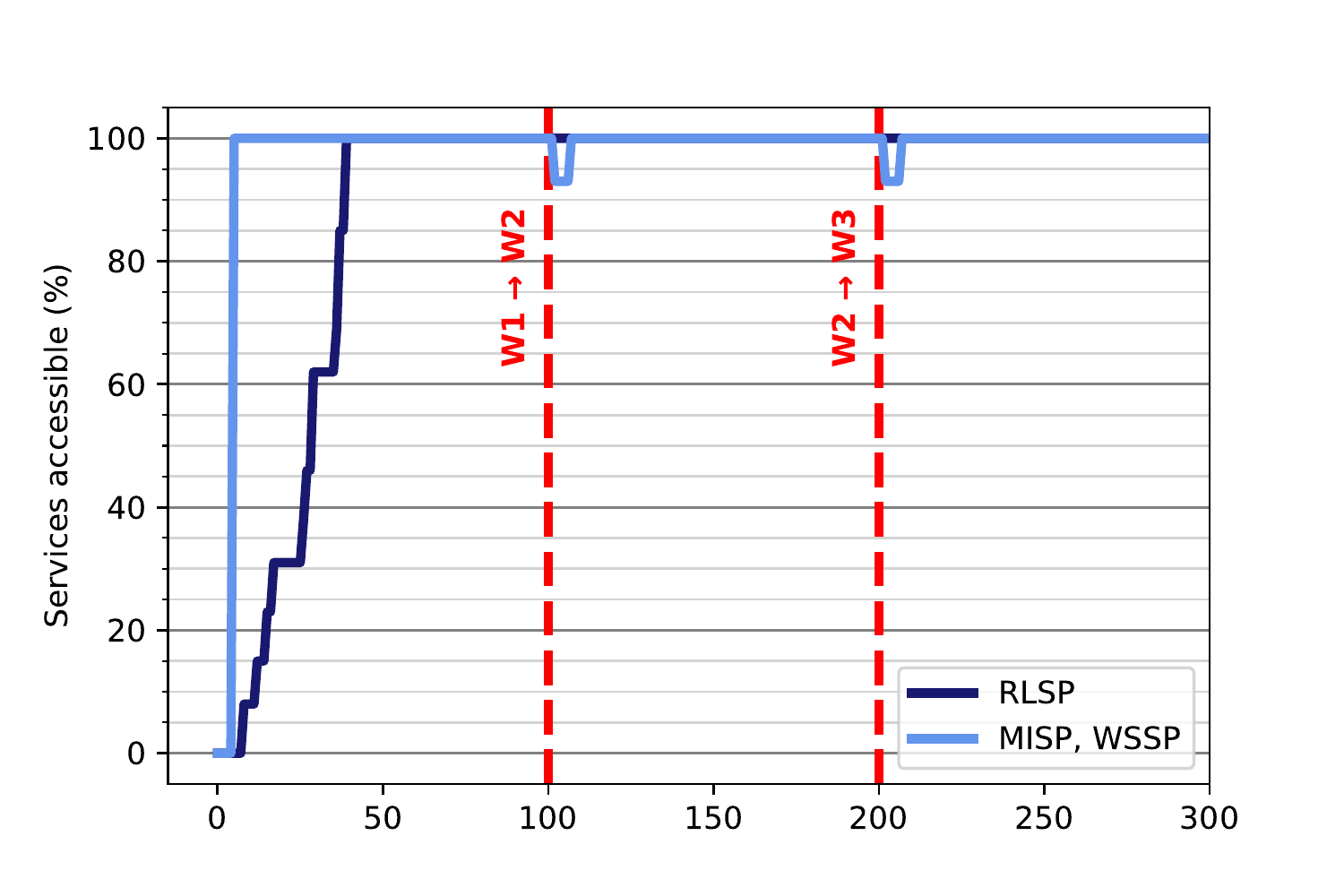}
    \caption{Service access variability during transitions between workloads}
    \label{fig:variability}
\end{figure}

 On the other hand, RLSP's placement solution remains the same. It means that in case the drone application service requirements are changed at run-time, WSSP and MISP algorithms will have a short time-span where some service chains will be temporarily inaccessible, while RLSP would have fewer or no disruptions. This is evident from Figure ~\ref{fig:variability} which shows the service access in the coverage area as the drone application transitions from $W1$ to $W2$ at timestamp 100, and from $W2$ to $W3$ at timestamp 200. The WSSP and MISP solutions make new deployments for $W2$ and $W3$ which makes 7\% service chains unavailable across multiple UE. This is restored after new deployments are completed at $E4$. The RLSP solution takes nearly 40 iterations to converge on its first solution. However, its solution is stable and requires no new deployments as for future workload transitions made by the application.

\par This demonstrates the advantage of RLSP to use past experience to reach placement solutions, i.e., the RL agent has sufficient prior knowledge of the system which helps it reach a better quality placement solution. The agent realizes that for stricter latencies, additional microservices need to be deployed, and is able to do so in advance. Thus, this behavior provides a stable deployment which will make relatively fewer changes as the application requirements change with time.

\subsection{Application deployment}
The application deployment criteria considers the number of microservices deployed, the cost incurred, the number of edge servers used and their load.
 
\subsubsection{Microservices deployed and deployment cost}
One of the aims of the placement algorithms is to reduce the deployment cost, while satisfying the application QoS and constraints. By optimizing the placement solution, the server load and the cost borne by infrastructure providers will be reduced. It must be noted here that WSSP has a relaxation since it does not support the service hardware requirements, microservice collocation, data locality constraints. For example, MISP and RLSP solutions deploy certain microservices on $E_{3}$ as they are bound by constraints. This can be seen in Figure~\ref{fig:microservicesDeployed} which shows the microservice counts per server for all the workloads and algorithms. 

\par The WSSP solution deploys a total 34, 38 and 42 microservices for the three workloads. MISP has a similar deployment but places 36, 40 and 44 microservices across the edge servers. The RLSP algorithm provides the minimum cost solution since it deploys only 23 microservices to provide 100\% service coverage for all three workloads. Thus, compared to WSSP and MISP, the RLSP solution deploys 32-45\% and 36-48\% fewer microservices respectively. This would lead to significant reduction in infrastructure cost. 


\par The difference in the number of microservices deployed across algorithms is because of the way they arrive at the final solution. WSSP makes one-time decisions considering all factors, and does not re-visit the result. MISP on the other hand deploy sufficient microservices to provide 100\% service chains across the coverage area, but it does not have a microservice eviction policy. Thus, both do not further try to minimize deployment cost while still meeting the application QoS requirements. The RLSP algorithm on the other hand, has an evict microservice action which enables it solution with minimum deployment cost. In the training process it was observed that RLSP first formed a complete but less optimized policy i.e. deployed more microservices, but provided 100\% service coverage. On further training, it also used its evict action to evict some microservices. This helped it converge to a final policy with less microservice deployment and 100\% coverage.  

\begin{figure*}[t]
    \centering
    
    \begin{subfigure}[b]{0.32\textwidth}
    \includegraphics[width=\textwidth]{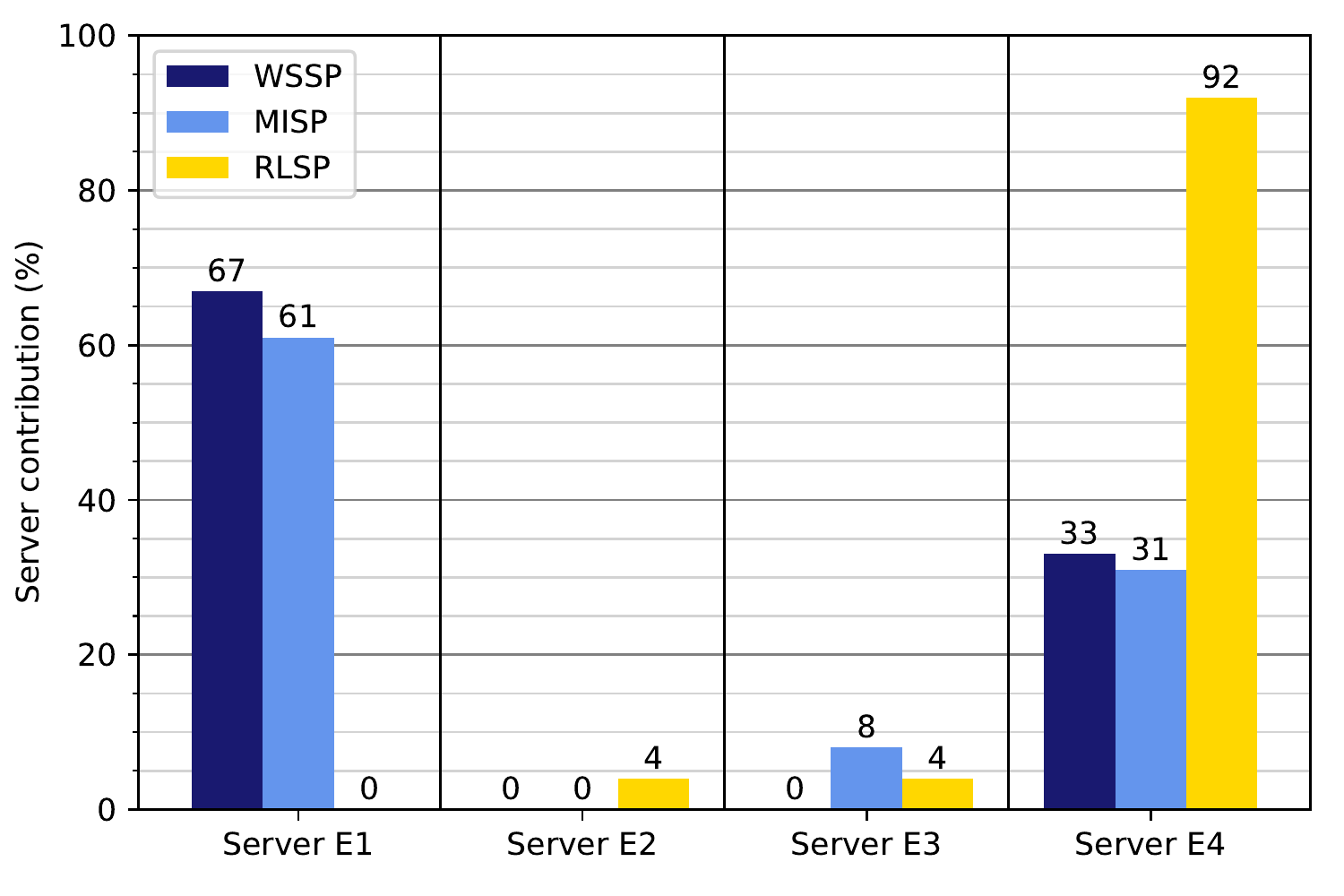}
    \caption{Relaxed workload}
    \label{fig:serverContribW1}
    \end{subfigure} \hfill
    \begin{subfigure}[b]{0.32\textwidth}
    \includegraphics[width=\textwidth]{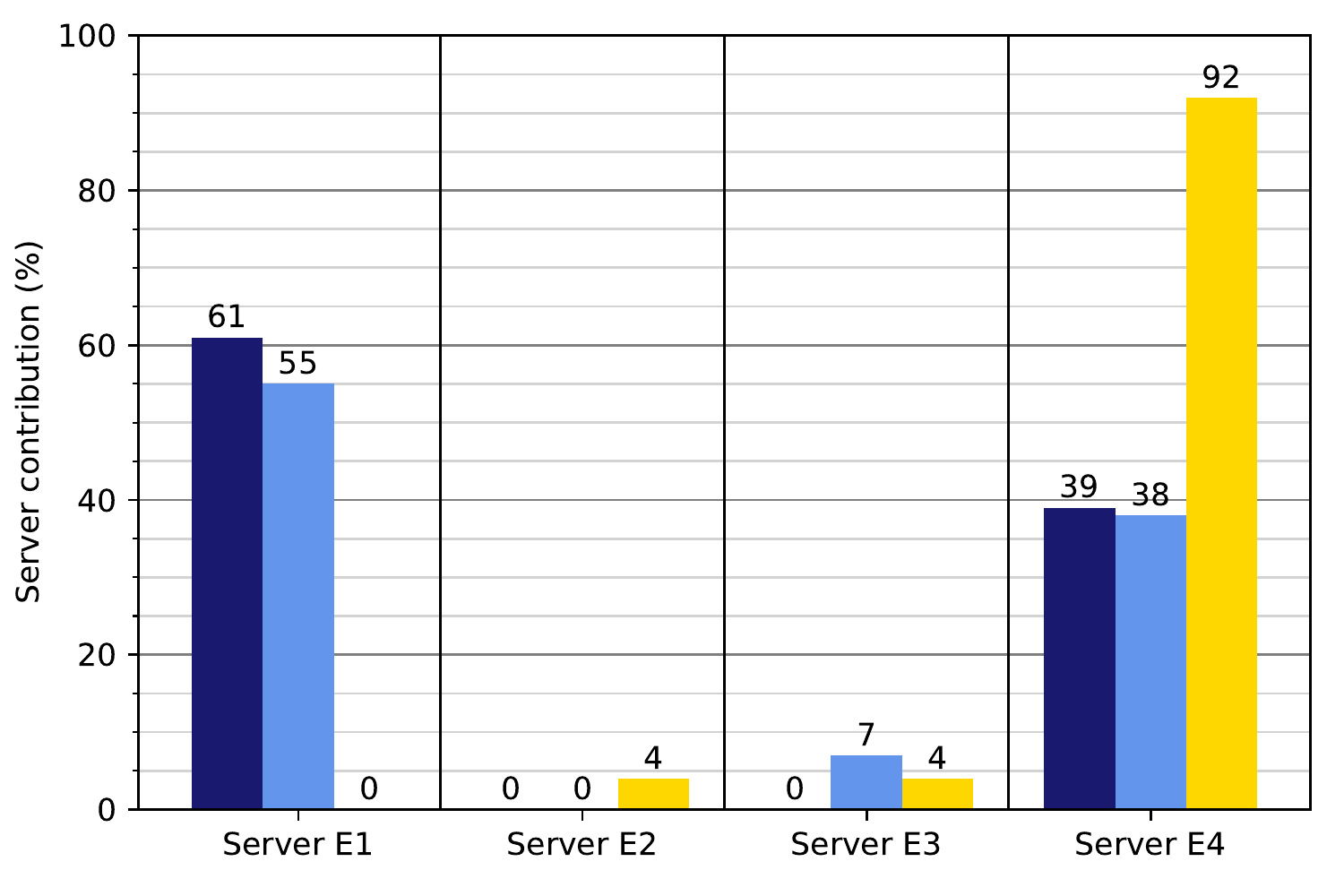}
    \caption{Moderate workload}
    \label{fig:serverContribW2}
    \end{subfigure} \hfill
    \begin{subfigure}[b]{0.32\textwidth}
    \includegraphics[width=\textwidth]{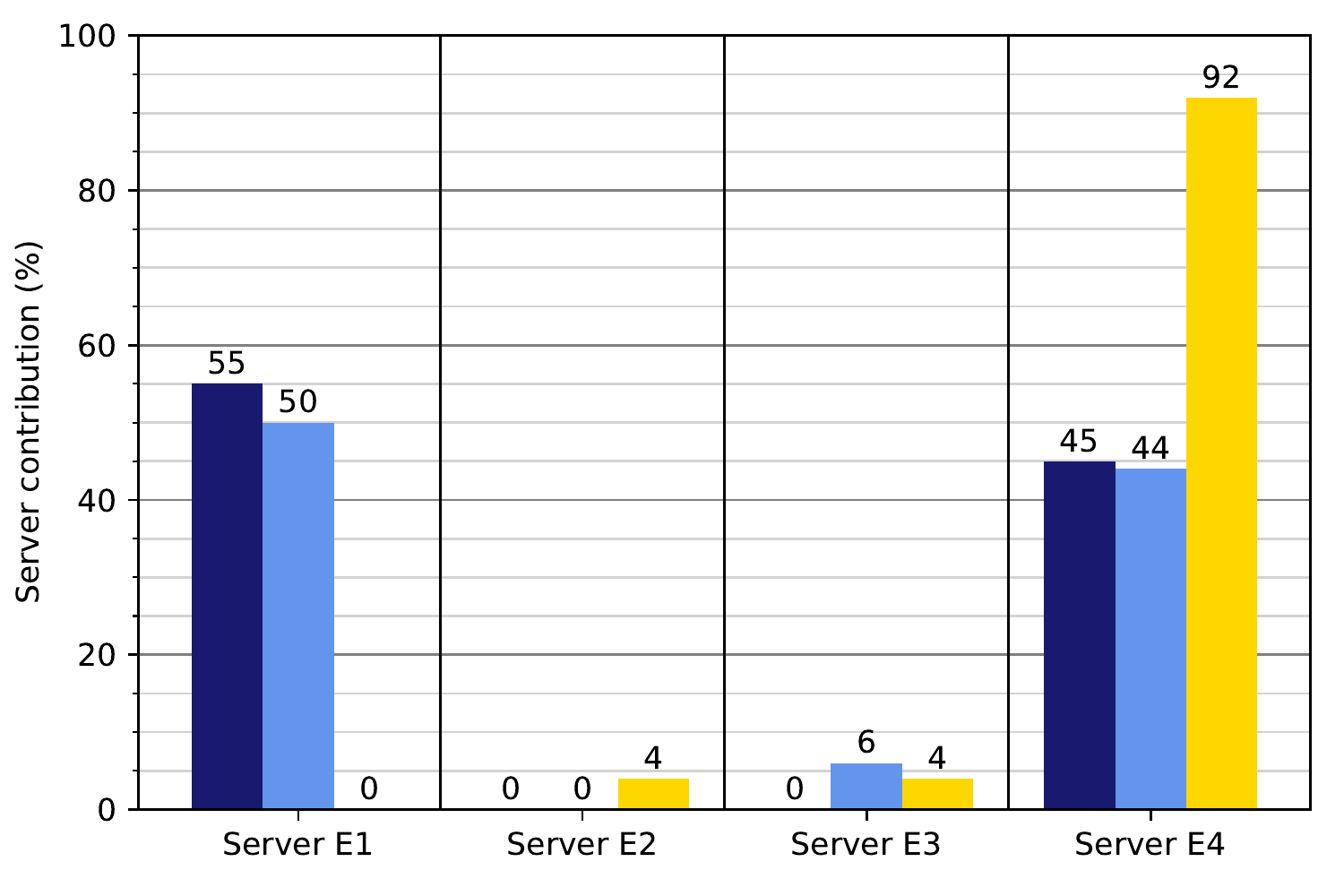}
    \caption{Ultra-low workload}
    \label{fig:serverContribW3}
    \end{subfigure}
    
    \caption{Server contribution in provisioning services}
    \label{fig:serverContrib}
\end{figure*}

\subsubsection{Edge Servers contribution and utilization}
All the three algorithms take into account the edge server site capacities and deploy microservices on edge servers as long as sufficient resources are available. In addition to the compute and storage resources, the algorithms also take into account the network link latency during communication from edge servers to UEs. Deploying microservices on edge servers having resources but large link-latency will not help the algorithms to provide a solution which satisfies the application QoS requirements. Thus, the algorithms take into account both the factors of resources and link latency while making placement decisions. 

\par Figure~\ref{fig:serverContrib} shows each server's contribution, as a percentage, in the microservice placement solutions produced by the three algorithms. It can be seen that most of the microservices are deployed among edge servers $E1$ and $E4$ since they are closer to many UEs. This provides ease of meeting QoS requirements through lower latency network links. The plot shows that server $E1$ contributes to 50-67\% of all microservices deployed by WSSP and MISP across the three workloads. Server $E4$ is also close to several UEs and some \textit{moderate} and \textit{relaxed} services were  provisioned through $E1$ for workload $W1$. However, as the QoS requirements became stricter in moderate $W2$ and ultra-low $W3$ workloads, WSSP and MISP deploy additional microservices on server $E4$ (as shown in Figure~\ref{fig:microservicesDeployed}). Across the three workloads, server $E4$ contributes to 31-45\% of the microservices deployed by WSSP and MISP. The RLSP algorithm finds the most optimal placement by placing 92\% of its total deployed microservices on $E4$. It also ensures that the deployment is within the server's resource limits.

\par In terms of server utilization, it is seen that edge servers $E2$ and $E3$ are not used at all by WSSP. They are utilized in a limited capacity by MISP and RLSP algorithms which place a few microservices on these edge servers to meet the application's constraints. Utilization of edge servers $E1$ and $E4$ is similar for WSSP and MISP algorithms across workloads. On the other hand, the RLSP algorithm utilizes server $E4$ to a considerable extent in all three workloads.

\subsection{Consistency of results}
To validate the deterministic nature of the algorithms, the experiments for each workload were re-run multiple times. The heuristic based approaches of WSSP and MISP - as implemented for our paper - are deterministic as they produced the same placement solution each time that they are given the same network conditions, application constraints and QoS requirements.  


The same cannot be said about RLSP. In contrast to WSSP and MISP algorithms which compute the solution through heuristic algorithms and mathematical solvers, the RLSP solution is obtained over multiple time-steps where it takes an action decision at each time-step, as described in Algorithm~\ref{alg:rl_env}. The RL agent in the RLSP algorithm takes actions based on its learning from the training stage and the given network deployment and application requirements. Based on its actions, the environment keeps generating the reward and a new state for the next time-step. Since these actions, reward values and environment states vary across timestamps, the path the RLSP algorithm takes to obtain the final converged placement solution is non-deterministic. In other words, although RLSP arrives at 100\% service accessibility result after several steps, the placement solution and the cumulative reward vary across runs. That being said, the placement solution for each run achieves all the placement objectives of service coverage, application constraints and QoS requirements. 


\begin{figure}[t]
    \centering
    \includegraphics[width=0.49\textwidth]{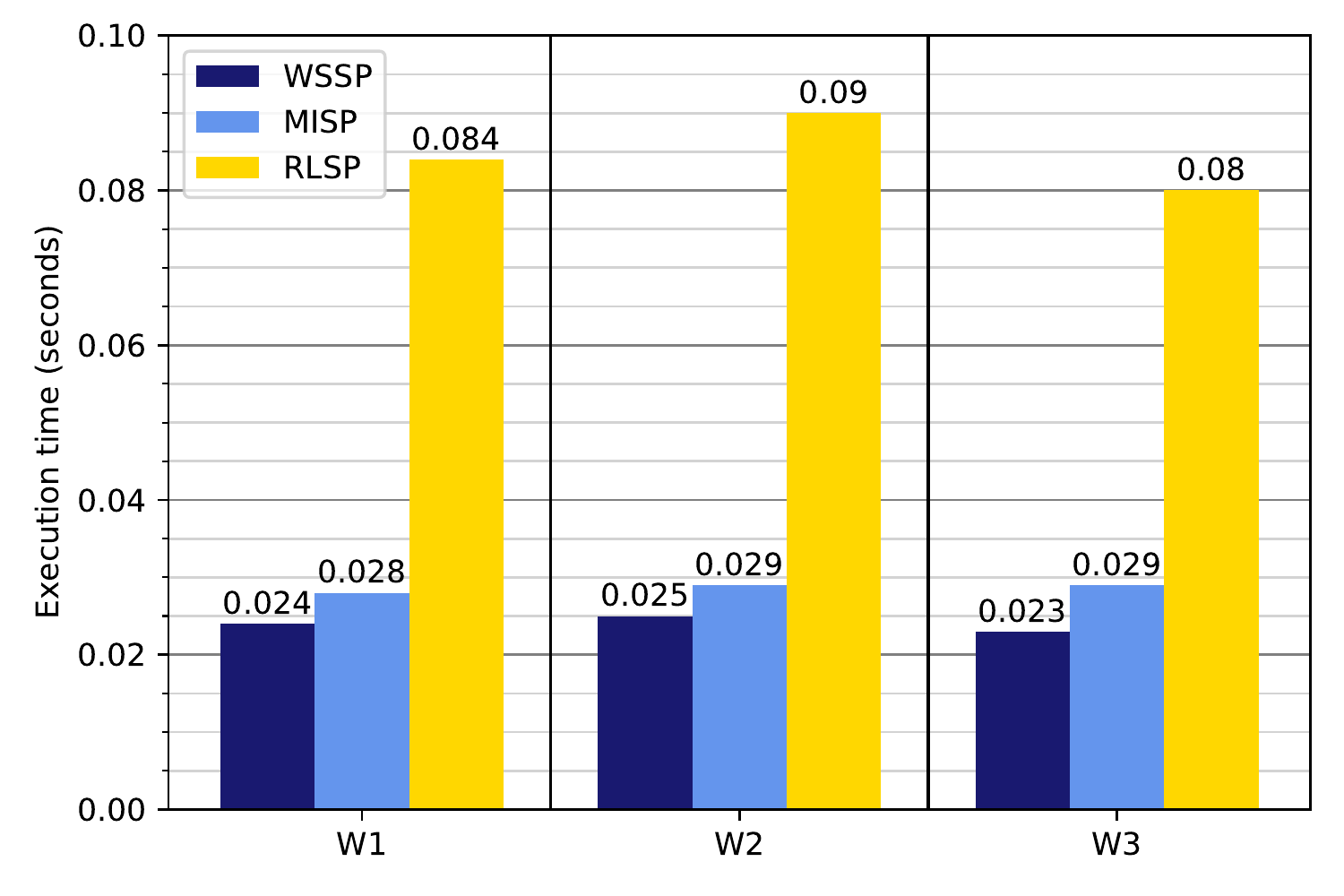}
    \caption{Execution time for the algorithms across the workloads} Hello
    \label{fig:execTime}
\end{figure}

\subsection{Algorithm execution time}
The time taken by the three algorithms to arrive at their final placement solutions is referred to as the algorithm execution time, shown in Figure~\ref{fig:execTime}. We see that WSSP is nearly 13.8\% faster than MISP since the former supports fewer constraints leading to saved computation time. The RLSP algorithm, as stated earlier, arrives at the solution after multiple time-steps. For the service placement problem, RLSP algorithm took nearly 40 iterations to converge at the final placement solution, and it spent  about 0.002 seconds at each step to to predict the next action. Thus, it spent about 0.08 seconds to find the placement solution for either of the three workloads.

\subsection{Comparative Analysis}\label{subsec:comp}

Our comparative evaluation showed that all three implemented algorithms were able to meet the latency requirements, but in different ways. The WSSP algorithm was not able to meet some constraints since they were not designed into the algorithm. 

Overall, the RLSP algorithm was able to provide better results, although it takes more time since it is based on reinforcement learning which converges after multiple iterations. An extensively trained RL agent is likely to experience much lesser variations in its solution (due to varying network conditions) compared to the other two algorithms. This provides stability in application deployment. It also deploys fewer microservices, leading to lower deployment costs, since it has an explicitly defined eviction policy. Although all three algorithms do provide good server utilization, it was seen that as the latency requirements got tighter, the RLSP algorithms showed better server utilization. However, due to its non-deterministic nature, it is not guaranteed to produce the same result after every run, although every result it produced did meet all the QoS requirements.

In conclusion, the three algorithms can be seen as a spectrum of dynamic placement solutions, ranging from the faster (WSSP) to slow (MISP) and slower (RLSP), although this is offset by solution quality (from constraint satisfaction viewpoint). Hence users can select the one most appropriate for their respective needs.

\section{Related work}\label{sec:related}
\subsection{Edge computing}

\par Dziyauddin et al.~\cite{dziyauddin2019computation} provides an overview of the architecture for vehicular edge computing, including  a review of computation offloading and content caching and delivery approaches for vehicular edge computing. Another work~\cite{elbamby2018proactive} investigates ultra-reliable and low-latency communication (URLLC) in fog networks, including a task distribution scheme with proactive caching of computing tasks at edge nodes. Efforts have also been made to build tools to test scheduling policies for edge computing~\cite{bauskar2020investigating}. 



\subsection{Dynamic service placement}
Overall, service placement for low latency and intensive applications bring conflicting cost functions including computation, communication and migration which need to be optimized, which we have demonstrated in this paper.

\par Virtual Reality (VR) and Augmented Reality (AR) applications have already adopted the edge computing paradigm to overcome the unpredictable latency of cloud-based processing. Wang et al.~\cite{wan2018application} places service entities for VR applications in the edge environment. Similarly, in VR group gaming, cloudlets are used~\cite{zhang2019dynamic}; while an algorithm for service placement for video analytics is presented in~\cite{farhadi2019service}.


\par Salaht et al.~\cite{salaht2020overview} present an overview of the several optimization metrics - mainly latency, cost and resource utilization - explored by recent works in service placement for fog and edge computing. Wang et al.~\cite{wang2018service} consider VR applications and formulate four cost types - activation, placement, proximity and collocation (resource contention among the services placed on the same edge cloud). We focus on the placement costs and also take into account proximity and collocation. Skarlat et al.~\cite{skarlat2017optimized}, like us, focus on QoS requirements and exploit the presence of edge computing to reduce QoS violations compared to a purely cloud-based approach. They also model applications as a combination of services and take into account the application requirements and the resources and latency links between the fogs (edges). However, unlike us, they do not take into account the utilization of resources in its placement and realistic network data such as communication delays.


Similar to ours, Ayoubi et al.~\cite{ayoubi2020autonomous} model equations for service latency, fog utilization, communication and computation costs as their key optimization objectives. He et al.~\cite{he2018s} categorize edge resources based on shareable (storage) and non-shareable (CPU cycles, bandwidth) resources for optimal provisioning of services and request scheduling. However, while deployment costs are considered by the other works, their placement strategies do \emph{not} support application specific constraints such as data locality and specialized hardware requirements which we have considered.

Mahmud et al.~\cite{mahmud2019quality} take a slightly different approach where their placement objective is based on user expectations and enhancement of Quality of Experience (QoE) with respect to access, service delivery and resource consumption. On similar lines,~\cite{ouyang2018follow} also lays emphasis on user mobility while optimizing edge service performance.


\par Salaht et al.~\cite{salaht2020overview} provide a clear service placement taxonomy based on four distinct aspects: control plane and coordination (centralized or distributed); placement characteristics (offline or online); nature of system (dynamic or not); mobility support (yes or no). In our work, we are mainly concerned with the second and third of these criteria.


\par \textbf{Heuristic based: } Skarlat et al.~\cite{skarlat2017optimized} design and implement a genetic algorithm to make service placement decisions. The genetic algorithm allows to investigate a large search space while providing viable solutions in polynomial time. The ITEM algorithm developed in~\cite{wang2018service} uses the graph-cut method to solve the combinatorial optimization problem based on a series of minimum $s-t$ cut instances. A similar approach is also taken in~\cite{zhang2019dynamic}. He et al.~\cite{he2018s} formulate service placement and request scheduling as an integer linear program aiming at serving maximum requests while minimizing the resource consumption. Similarly, Farhadi et al.~\cite{farhadi2019service} take a joint approach in optimizing service placement and request scheduling by combining greedy heuristic and shadow request scheduling techniques.


\par \textbf{Learning based: } Ayoubi et al.~\cite{ayoubi2020autonomous} propose a four-stage MADE strategy which consists of monitoring, analysis, decision making and execution phases. Ouyang et al.~\cite{ouyang2019adaptive} take an adaptive user-centric approach towards service placement which also considers user-mobility and user's preferences, and which uses an online learning based algorithm. Brandherm et al.~\cite{brandherm2019learning} uses a model-free Q-learning based approach for service migration. Finally, Gao et al.~\cite{gao2019winning} delve into preserving application QoS in the mobile edge computing framework by also laying emphasis on the network factors, since both access network and edge nodes are vulnerable to congestion.



\par None of the above works provide any sort of detailed comparison among specific algorithms across the spectrum encompassing heuristics (simpler and more complicated) and learning-based algorithms to provide practitioners a possible range of solutions with various tradeoffs.
\section{Conclusion and Future work}\label{sec:conc}
In this paper, we have addressed the well-known dynamic service placement problem. In contrast to earlier works that usually present only a main single solution to the problem from a single perspective (albeit with variations), our paper proposes and compares three diverse algorithms to resolve this problem. These algorithms cover the spectrum from simpler heuristics to a more involved learning-based approach. In so doing, we have presented the algorithms, as well as a detailed comparative evaluation on a realistic drone swarm application. While the results of our comparative evaluation demonstrate that the learning-based approach does provide better results, we have also shown that the simpler heuristics are not to be ruled out, especially when a trade-off such as execution time is considered. Hence the work in our paper can be considered as - to the best of our knowledge - the first such detailed comparative study of dynamic placement algorithms along with recommendations to practitioners on the best scenarios in which each of them can be used. 

Another novel aspect of our paper has been to incorporate the intricacies of the underlying mobile network infrastructure while developing our algorithms, something that is also - to the best of our knowledge - the first such detailed treatment of this problem. We hope that this will give impetus to further research on dynamic service placement approaches that incorporate the underlying mobile network infrastructure. 


Hence our future work will involve not only the above incorporation of mobile network infrastructure into dynamic service placement approaches, it will also involve the following: (1) development and evaluation of components that actually implement microservice migration across edge servers (perhaps either on a network emulator or a real 5G testbed); (2) enrichment of microservice migration via integration of service mesh technology~\cite{li2019service} for dynamic management of microservice traffic, especially to demonstrate microservice migration on a large-scale; and (3) investigation of how our algorithms can be deployed and evaluated on real-life examples (perhaps such as those described in~\cite{gan2019open}) on a large scale.
\begin{acks}

We wish to thank Ajay Kattepur for assistance on RLSP, as well as our respective colleagues in Erisson Research for their feedback.
\end{acks}

\bibliographystyle{ACM-Reference-Format}
\bibliography{sample-base}


\begin{thebibliography}{30}


\ifx \showCODEN    \undefined \def \showCODEN     #1{\unskip}     \fi
\ifx \showDOI      \undefined \def \showDOI       #1{#1}\fi
\ifx \showISBNx    \undefined \def \showISBNx     #1{\unskip}     \fi
\ifx \showISBNxiii \undefined \def \showISBNxiii  #1{\unskip}     \fi
\ifx \showISSN     \undefined \def \showISSN      #1{\unskip}     \fi
\ifx \showLCCN     \undefined \def \showLCCN      #1{\unskip}     \fi
\ifx \shownote     \undefined \def \shownote      #1{#1}          \fi
\ifx \showarticletitle \undefined \def \showarticletitle #1{#1}   \fi
\ifx \showURL      \undefined \def \showURL       {\relax}        \fi
\providecommand\bibfield[2]{#2}
\providecommand\bibinfo[2]{#2}
\providecommand\natexlab[1]{#1}
\providecommand\showeprint[2][]{arXiv:#2}

\bibitem[\protect\citeauthoryear{Ayoubi, Ramezanpour, and Khorsand}{Ayoubi
  et~al\mbox{.}}{2020}]%
        {ayoubi2020autonomous}
\bibfield{author}{\bibinfo{person}{Masoumeh Ayoubi},
  \bibinfo{person}{Mohammadreza Ramezanpour}, {and} \bibinfo{person}{Reihaneh
  Khorsand}.} \bibinfo{year}{2020}\natexlab{}.
\newblock \showarticletitle{An autonomous IoT service placement methodology in
  fog computing}.
\newblock \bibinfo{journal}{\emph{Software: Practice and Experience}}
  (\bibinfo{year}{2020}).
\newblock


\bibitem[\protect\citeauthoryear{Bahreini and Grosu}{Bahreini and
  Grosu}{2017}]%
        {bahreini2017efficient}
\bibfield{author}{\bibinfo{person}{Tayebeh Bahreini} {and}
  \bibinfo{person}{Daniel Grosu}.} \bibinfo{year}{2017}\natexlab{}.
\newblock \showarticletitle{Efficient placement of multi-component applications
  in edge computing systems}. In \bibinfo{booktitle}{\emph{Proceedings of the
  Second ACM/IEEE Symposium on Edge Computing}}. \bibinfo{pages}{1--11}.
\newblock


\bibitem[\protect\citeauthoryear{Bauskar, Da~Silva, Lebre, Mommessin, Neyron,
  Ngoko, Ricordel, Trystram, and van Kempen}{Bauskar et~al\mbox{.}}{2020}]%
        {bauskar2020investigating}
\bibfield{author}{\bibinfo{person}{Adwait Bauskar}, \bibinfo{person}{Anderson
  Da~Silva}, \bibinfo{person}{Adrien Lebre}, \bibinfo{person}{Clement
  Mommessin}, \bibinfo{person}{Pierre Neyron}, \bibinfo{person}{Yanik Ngoko},
  \bibinfo{person}{Yoann Ricordel}, \bibinfo{person}{Denis Trystram}, {and}
  \bibinfo{person}{Alexandre van Kempen}.} \bibinfo{year}{2020}\natexlab{}.
\newblock \emph{\bibinfo{title}{Investigating Placement Challenges in Edge
  Infrastructures through a Common Simulator (extended version)}}.
\newblock \bibinfo{thesistype}{Ph.D. Dissertation}. \bibinfo{school}{INRIA}.
\newblock


\bibitem[\protect\citeauthoryear{Bor-Yaliniz, Salem, Senerath, and
  Yanikomeroglu}{Bor-Yaliniz et~al\mbox{.}}{2019}]%
        {bor20195g}
\bibfield{author}{\bibinfo{person}{Irem Bor-Yaliniz}, \bibinfo{person}{Mohamed
  Salem}, \bibinfo{person}{Gamini Senerath}, {and} \bibinfo{person}{Halim
  Yanikomeroglu}.} \bibinfo{year}{2019}\natexlab{}.
\newblock \showarticletitle{Is 5G ready for drones: A look into contemporary
  and prospective wireless networks from a standardization perspective}.
\newblock \bibinfo{journal}{\emph{IEEE Wireless Communications}}
  \bibinfo{volume}{26}, \bibinfo{number}{1} (\bibinfo{year}{2019}),
  \bibinfo{pages}{18--27}.
\newblock


\bibitem[\protect\citeauthoryear{Brandherm, Wang, and
  M{\"u}hlh{\"a}user}{Brandherm et~al\mbox{.}}{2019}]%
        {brandherm2019learning}
\bibfield{author}{\bibinfo{person}{Florian Brandherm}, \bibinfo{person}{Lin
  Wang}, {and} \bibinfo{person}{Max M{\"u}hlh{\"a}user}.}
  \bibinfo{year}{2019}\natexlab{}.
\newblock \showarticletitle{A learning-based framework for optimizing service
  migration in mobile edge clouds}. In \bibinfo{booktitle}{\emph{Proceedings of
  the 2nd International Workshop on Edge Systems, Analytics and Networking}}.
  \bibinfo{pages}{12--17}.
\newblock


\bibitem[\protect\citeauthoryear{Brockman, Cheung, Pettersson, Schneider,
  Schulman, Tang, and Zaremba}{Brockman et~al\mbox{.}}{2016}]%
        {brockman2016openai}
\bibfield{author}{\bibinfo{person}{Greg Brockman}, \bibinfo{person}{Vicki
  Cheung}, \bibinfo{person}{Ludwig Pettersson}, \bibinfo{person}{Jonas
  Schneider}, \bibinfo{person}{John Schulman}, \bibinfo{person}{Jie Tang},
  {and} \bibinfo{person}{Wojciech Zaremba}.} \bibinfo{year}{2016}\natexlab{}.
\newblock \showarticletitle{Openai gym}.
\newblock \bibinfo{journal}{\emph{arXiv preprint arXiv:1606.01540}}
  (\bibinfo{year}{2016}).
\newblock


\bibitem[\protect\citeauthoryear{Dziyauddin, Niyato, Luong, Izhar, Hadhari, and
  Daud}{Dziyauddin et~al\mbox{.}}{2019}]%
        {dziyauddin2019computation}
\bibfield{author}{\bibinfo{person}{Rudzidatul~Akmam Dziyauddin},
  \bibinfo{person}{Dusit Niyato}, \bibinfo{person}{Nguyen~Cong Luong},
  \bibinfo{person}{Mohd Azri~Mohd Izhar}, \bibinfo{person}{Marwan Hadhari},
  {and} \bibinfo{person}{Salwani Daud}.} \bibinfo{year}{2019}\natexlab{}.
\newblock \showarticletitle{Computation offloading and content caching delivery
  in vehicular edge computing: A survey}.
\newblock \bibinfo{journal}{\emph{arXiv preprint arXiv:1912.07803}}
  (\bibinfo{year}{2019}).
\newblock


\bibitem[\protect\citeauthoryear{Elbamby, Bennis, Saad, Latva-Aho, and
  Hong}{Elbamby et~al\mbox{.}}{2018}]%
        {elbamby2018proactive}
\bibfield{author}{\bibinfo{person}{Mohammed~S Elbamby}, \bibinfo{person}{Mehdi
  Bennis}, \bibinfo{person}{Walid Saad}, \bibinfo{person}{Matti Latva-Aho},
  {and} \bibinfo{person}{Choong~Seon Hong}.} \bibinfo{year}{2018}\natexlab{}.
\newblock \showarticletitle{Proactive edge computing in fog networks with
  latency and reliability guarantees}.
\newblock \bibinfo{journal}{\emph{EURASIP Journal on Wireless Communications
  and Networking}} \bibinfo{volume}{2018}, \bibinfo{number}{1}
  (\bibinfo{year}{2018}), \bibinfo{pages}{1--13}.
\newblock


\bibitem[\protect\citeauthoryear{Fan and Li}{Fan and Li}{2020}]%
        {Fan_2020}
\bibfield{author}{\bibinfo{person}{Chen Fan} {and} \bibinfo{person}{Li Li}.}
  \bibinfo{year}{2020}\natexlab{}.
\newblock \showarticletitle{Service Migration in Mobile Edge Computing Based on
  Reinforcement Learning}.
\newblock \bibinfo{journal}{\emph{Journal of Physics: Conference Series}}
  \bibinfo{volume}{1584} (\bibinfo{date}{jul} \bibinfo{year}{2020}),
  \bibinfo{pages}{012058}.
\newblock
\urldef\tempurl%
\url{https://doi.org/10.1088/1742-6596/1584/1/012058}
\showDOI{\tempurl}


\bibitem[\protect\citeauthoryear{Farhadi, Mehmeti, He, La~Porta, Khamfroush,
  Wang, and Chan}{Farhadi et~al\mbox{.}}{2019}]%
        {farhadi2019service}
\bibfield{author}{\bibinfo{person}{Vajiheh Farhadi}, \bibinfo{person}{Fidan
  Mehmeti}, \bibinfo{person}{Ting He}, \bibinfo{person}{Tom La~Porta},
  \bibinfo{person}{Hana Khamfroush}, \bibinfo{person}{Shiqiang Wang}, {and}
  \bibinfo{person}{Kevin~S Chan}.} \bibinfo{year}{2019}\natexlab{}.
\newblock \showarticletitle{Service placement and request scheduling for
  data-intensive applications in edge clouds}. In
  \bibinfo{booktitle}{\emph{IEEE INFOCOM 2019-IEEE Conference on Computer
  Communications}}. IEEE, \bibinfo{pages}{1279--1287}.
\newblock


\bibitem[\protect\citeauthoryear{Gan, Zhang, Cheng, Shetty, Rathi, Katarki,
  Bruno, Hu, Ritchken, Jackson, et~al\mbox{.}}{Gan et~al\mbox{.}}{2019}]%
        {gan2019open}
\bibfield{author}{\bibinfo{person}{Yu Gan}, \bibinfo{person}{Yanqi Zhang},
  \bibinfo{person}{Dailun Cheng}, \bibinfo{person}{Ankitha Shetty},
  \bibinfo{person}{Priyal Rathi}, \bibinfo{person}{Nayan Katarki},
  \bibinfo{person}{Ariana Bruno}, \bibinfo{person}{Justin Hu},
  \bibinfo{person}{Brian Ritchken}, \bibinfo{person}{Brendon Jackson},
  {et~al\mbox{.}}} \bibinfo{year}{2019}\natexlab{}.
\newblock \showarticletitle{An open-source benchmark suite for microservices
  and their hardware-software implications for cloud \& edge systems}. In
  \bibinfo{booktitle}{\emph{Proceedings of the Twenty-Fourth International
  Conference on Architectural Support for Programming Languages and Operating
  Systems}}. \bibinfo{pages}{3--18}.
\newblock


\bibitem[\protect\citeauthoryear{Gao, Zhou, Liu, and Xu}{Gao
  et~al\mbox{.}}{2019}]%
        {gao2019winning}
\bibfield{author}{\bibinfo{person}{Bin Gao}, \bibinfo{person}{Zhi Zhou},
  \bibinfo{person}{Fangming Liu}, {and} \bibinfo{person}{Fei Xu}.}
  \bibinfo{year}{2019}\natexlab{}.
\newblock \showarticletitle{Winning at the starting line: Joint network
  selection and service placement for mobile edge computing}. In
  \bibinfo{booktitle}{\emph{IEEE INFOCOM 2019-IEEE Conference on Computer
  Communications}}. IEEE, \bibinfo{pages}{1459--1467}.
\newblock


\bibitem[\protect\citeauthoryear{Hassin and Levin}{Hassin and Levin}{2005}]%
        {hassin2005better}
\bibfield{author}{\bibinfo{person}{Refael Hassin} {and} \bibinfo{person}{Asaf
  Levin}.} \bibinfo{year}{2005}\natexlab{}.
\newblock \showarticletitle{A better-than-greedy approximation algorithm for
  the minimum set cover problem}.
\newblock \bibinfo{journal}{\emph{SIAM J. Comput.}} \bibinfo{volume}{35},
  \bibinfo{number}{1} (\bibinfo{year}{2005}), \bibinfo{pages}{189--200}.
\newblock


\bibitem[\protect\citeauthoryear{He, Khamfroush, Wang, La~Porta, and Stein}{He
  et~al\mbox{.}}{2018}]%
        {he2018s}
\bibfield{author}{\bibinfo{person}{Ting He}, \bibinfo{person}{Hana Khamfroush},
  \bibinfo{person}{Shiqiang Wang}, \bibinfo{person}{Tom La~Porta}, {and}
  \bibinfo{person}{Sebastian Stein}.} \bibinfo{year}{2018}\natexlab{}.
\newblock \showarticletitle{It's hard to share: Joint service placement and
  request scheduling in edge clouds with sharable and non-sharable resources}.
  In \bibinfo{booktitle}{\emph{2018 IEEE 38th International Conference on
  Distributed Computing Systems (ICDCS)}}. IEEE, \bibinfo{pages}{365--375}.
\newblock


\bibitem[\protect\citeauthoryear{Li, Lemieux, Gao, Zhao, and Han}{Li
  et~al\mbox{.}}{2019}]%
        {li2019service}
\bibfield{author}{\bibinfo{person}{Wubin Li}, \bibinfo{person}{Yves Lemieux},
  \bibinfo{person}{Jing Gao}, \bibinfo{person}{Zhuofeng Zhao}, {and}
  \bibinfo{person}{Yanbo Han}.} \bibinfo{year}{2019}\natexlab{}.
\newblock \showarticletitle{Service mesh: Challenges, state of the art, and
  future research opportunities}. In \bibinfo{booktitle}{\emph{2019 IEEE
  International Conference on Service-Oriented System Engineering (SOSE)}}.
  IEEE, \bibinfo{pages}{122--1225}.
\newblock


\bibitem[\protect\citeauthoryear{L{\'o}pez-Pires, Bar{\'a}n, Ben{\'\i}tez,
  Zalimben, and Amarilla}{L{\'o}pez-Pires et~al\mbox{.}}{2018}]%
        {lopez2018virtual}
\bibfield{author}{\bibinfo{person}{Fabio L{\'o}pez-Pires},
  \bibinfo{person}{Benjam{\'\i}n Bar{\'a}n}, \bibinfo{person}{Leonardo
  Ben{\'\i}tez}, \bibinfo{person}{Sa{\'u}l Zalimben}, {and}
  \bibinfo{person}{Augusto Amarilla}.} \bibinfo{year}{2018}\natexlab{}.
\newblock \showarticletitle{Virtual machine placement for elastic
  infrastructures in overbooked cloud computing datacenters under uncertainty}.
\newblock \bibinfo{journal}{\emph{Future Generation Computer Systems}}
  \bibinfo{volume}{79} (\bibinfo{year}{2018}), \bibinfo{pages}{830--848}.
\newblock


\bibitem[\protect\citeauthoryear{Mahmud, Srirama, Ramamohanarao, and
  Buyya}{Mahmud et~al\mbox{.}}{2019}]%
        {mahmud2019quality}
\bibfield{author}{\bibinfo{person}{Redowan Mahmud},
  \bibinfo{person}{Satish~Narayana Srirama}, \bibinfo{person}{Kotagiri
  Ramamohanarao}, {and} \bibinfo{person}{Rajkumar Buyya}.}
  \bibinfo{year}{2019}\natexlab{}.
\newblock \showarticletitle{Quality of Experience (QoE)-aware placement of
  applications in Fog computing environments}.
\newblock \bibinfo{journal}{\emph{J. Parallel and Distrib. Comput.}}
  \bibinfo{volume}{132} (\bibinfo{year}{2019}), \bibinfo{pages}{190--203}.
\newblock


\bibitem[\protect\citeauthoryear{Marler and Arora}{Marler and Arora}{2004}]%
        {marler2004survey}
\bibfield{author}{\bibinfo{person}{R~Timothy Marler} {and}
  \bibinfo{person}{Jasbir~S Arora}.} \bibinfo{year}{2004}\natexlab{}.
\newblock \showarticletitle{Survey of multi-objective optimization methods for
  engineering}.
\newblock \bibinfo{journal}{\emph{Structural and multidisciplinary
  optimization}} \bibinfo{volume}{26}, \bibinfo{number}{6}
  (\bibinfo{year}{2004}), \bibinfo{pages}{369--395}.
\newblock


\bibitem[\protect\citeauthoryear{Mitchell}{Mitchell}{1997}]%
        {mitchell1997machine}
\bibfield{author}{\bibinfo{person}{TM Mitchell}.}
  \bibinfo{year}{1997}\natexlab{}.
\newblock \showarticletitle{Machine Learning, McGraw-Hill Higher Education}.
\newblock \bibinfo{journal}{\emph{New York}} (\bibinfo{year}{1997}).
\newblock


\bibitem[\protect\citeauthoryear{Ouyang, Li, Chen, Zhou, and Tang}{Ouyang
  et~al\mbox{.}}{2019}]%
        {ouyang2019adaptive}
\bibfield{author}{\bibinfo{person}{Tao Ouyang}, \bibinfo{person}{Rui Li},
  \bibinfo{person}{Xu Chen}, \bibinfo{person}{Zhi Zhou}, {and}
  \bibinfo{person}{Xin Tang}.} \bibinfo{year}{2019}\natexlab{}.
\newblock \showarticletitle{Adaptive user-managed service placement for mobile
  edge computing: An online learning approach}. In
  \bibinfo{booktitle}{\emph{IEEE INFOCOM 2019-IEEE Conference on Computer
  Communications}}. IEEE, \bibinfo{pages}{1468--1476}.
\newblock


\bibitem[\protect\citeauthoryear{Ouyang, Zhou, and Chen}{Ouyang
  et~al\mbox{.}}{2018}]%
        {ouyang2018follow}
\bibfield{author}{\bibinfo{person}{Tao Ouyang}, \bibinfo{person}{Zhi Zhou},
  {and} \bibinfo{person}{Xu Chen}.} \bibinfo{year}{2018}\natexlab{}.
\newblock \showarticletitle{Follow me at the edge: Mobility-aware dynamic
  service placement for mobile edge computing}.
\newblock \bibinfo{journal}{\emph{IEEE Journal on Selected Areas in
  Communications}} \bibinfo{volume}{36}, \bibinfo{number}{10}
  (\bibinfo{year}{2018}), \bibinfo{pages}{2333--2345}.
\newblock


\bibitem[\protect\citeauthoryear{Raffin, Hill, Ernestus, Gleave, Kanervisto,
  and Dormann}{Raffin et~al\mbox{.}}{2019}]%
        {stable-baselines3}
\bibfield{author}{\bibinfo{person}{Antonin Raffin}, \bibinfo{person}{Ashley
  Hill}, \bibinfo{person}{Maximilian Ernestus}, \bibinfo{person}{Adam Gleave},
  \bibinfo{person}{Anssi Kanervisto}, {and} \bibinfo{person}{Noah Dormann}.}
  \bibinfo{year}{2019}\natexlab{}.
\newblock \bibinfo{title}{Stable Baselines3}.
\newblock
  \bibinfo{howpublished}{\url{https://github.com/DLR-RM/stable-baselines3}}.
\newblock


\bibitem[\protect\citeauthoryear{Salaht, Desprez, and Lebre}{Salaht
  et~al\mbox{.}}{2020}]%
        {salaht2020overview}
\bibfield{author}{\bibinfo{person}{Farah~Ait Salaht},
  \bibinfo{person}{Fr{\'e}d{\'e}ric Desprez}, {and} \bibinfo{person}{Adrien
  Lebre}.} \bibinfo{year}{2020}\natexlab{}.
\newblock \showarticletitle{An overview of service placement problem in Fog and
  Edge Computing}.
\newblock \bibinfo{journal}{\emph{ACM Computing Surveys (CSUR)}}
  \bibinfo{volume}{53}, \bibinfo{number}{3} (\bibinfo{year}{2020}),
  \bibinfo{pages}{1--35}.
\newblock


\bibitem[\protect\citeauthoryear{Schulman, Wolski, Dhariwal, Radford, and
  Klimov}{Schulman et~al\mbox{.}}{2017}]%
        {schulman2017proximal}
\bibfield{author}{\bibinfo{person}{John Schulman}, \bibinfo{person}{Filip
  Wolski}, \bibinfo{person}{Prafulla Dhariwal}, \bibinfo{person}{Alec Radford},
  {and} \bibinfo{person}{Oleg Klimov}.} \bibinfo{year}{2017}\natexlab{}.
\newblock \showarticletitle{Proximal policy optimization algorithms}.
\newblock \bibinfo{journal}{\emph{arXiv preprint arXiv:1707.06347}}
  (\bibinfo{year}{2017}).
\newblock


\bibitem[\protect\citeauthoryear{Sedaghat, Hern{\'a}ndez-Rodriguez, and
  Elmroth}{Sedaghat et~al\mbox{.}}{2016}]%
        {sedaghat2016decentralized}
\bibfield{author}{\bibinfo{person}{Mina Sedaghat}, \bibinfo{person}{Francisco
  Hern{\'a}ndez-Rodriguez}, {and} \bibinfo{person}{Erik Elmroth}.}
  \bibinfo{year}{2016}\natexlab{}.
\newblock \showarticletitle{Decentralized cloud datacenter reconsolidation
  through emergent and topology-aware behavior}.
\newblock \bibinfo{journal}{\emph{Future Generation Computer Systems}}
  \bibinfo{volume}{56} (\bibinfo{year}{2016}), \bibinfo{pages}{51--63}.
\newblock


\bibitem[\protect\citeauthoryear{Skarlat, Nardelli, Schulte, Borkowski, and
  Leitner}{Skarlat et~al\mbox{.}}{2017}]%
        {skarlat2017optimized}
\bibfield{author}{\bibinfo{person}{Olena Skarlat}, \bibinfo{person}{Matteo
  Nardelli}, \bibinfo{person}{Stefan Schulte}, \bibinfo{person}{Michael
  Borkowski}, {and} \bibinfo{person}{Philipp Leitner}.}
  \bibinfo{year}{2017}\natexlab{}.
\newblock \showarticletitle{Optimized IoT service placement in the fog}.
\newblock \bibinfo{journal}{\emph{Service Oriented Computing and Applications}}
  \bibinfo{volume}{11}, \bibinfo{number}{4} (\bibinfo{year}{2017}),
  \bibinfo{pages}{427--443}.
\newblock


\bibitem[\protect\citeauthoryear{Sv{\"a}rd, Li, Wadbro, Tordsson, and
  Elmroth}{Sv{\"a}rd et~al\mbox{.}}{2015}]%
        {svard2015continuous}
\bibfield{author}{\bibinfo{person}{Petter Sv{\"a}rd}, \bibinfo{person}{Wubin
  Li}, \bibinfo{person}{Eddie Wadbro}, \bibinfo{person}{Johan Tordsson}, {and}
  \bibinfo{person}{Erik Elmroth}.} \bibinfo{year}{2015}\natexlab{}.
\newblock \showarticletitle{Continuous datacenter consolidation}. In
  \bibinfo{booktitle}{\emph{2015 IEEE 7th International Conference on Cloud
  Computing Technology and Science (CloudCom)}}. IEEE,
  \bibinfo{pages}{387--396}.
\newblock


\bibitem[\protect\citeauthoryear{Wan, Guan, Wang, Bai, and Choi}{Wan
  et~al\mbox{.}}{2018}]%
        {wan2018application}
\bibfield{author}{\bibinfo{person}{Xili Wan}, \bibinfo{person}{Xinjie Guan},
  \bibinfo{person}{Tianjing Wang}, \bibinfo{person}{Guangwei Bai}, {and}
  \bibinfo{person}{Baek-Yong Choi}.} \bibinfo{year}{2018}\natexlab{}.
\newblock \showarticletitle{Application deployment using Microservice and
  Docker containers: Framework and optimization}.
\newblock \bibinfo{journal}{\emph{Journal of Network and Computer
  Applications}}  \bibinfo{volume}{119} (\bibinfo{year}{2018}),
  \bibinfo{pages}{97--109}.
\newblock


\bibitem[\protect\citeauthoryear{Wang, Jiao, He, Li, and
  M{\"u}hlh{\"a}user}{Wang et~al\mbox{.}}{2018}]%
        {wang2018service}
\bibfield{author}{\bibinfo{person}{Lin Wang}, \bibinfo{person}{Lei Jiao},
  \bibinfo{person}{Ting He}, \bibinfo{person}{Jun Li}, {and}
  \bibinfo{person}{Max M{\"u}hlh{\"a}user}.} \bibinfo{year}{2018}\natexlab{}.
\newblock \showarticletitle{Service entity placement for social virtual reality
  applications in edge computing}. In \bibinfo{booktitle}{\emph{IEEE INFOCOM
  2018-IEEE Conference on Computer Communications}}. IEEE,
  \bibinfo{pages}{468--476}.
\newblock


\bibitem[\protect\citeauthoryear{Zhang, Jiao, Yan, and Lin}{Zhang
  et~al\mbox{.}}{2019}]%
        {zhang2019dynamic}
\bibfield{author}{\bibinfo{person}{Yuan Zhang}, \bibinfo{person}{Lei Jiao},
  \bibinfo{person}{Jinyao Yan}, {and} \bibinfo{person}{Xiaojun Lin}.}
  \bibinfo{year}{2019}\natexlab{}.
\newblock \showarticletitle{Dynamic service placement for virtual reality group
  gaming on mobile edge cloudlets}.
\newblock \bibinfo{journal}{\emph{IEEE Journal on Selected Areas in
  Communications}} \bibinfo{volume}{37}, \bibinfo{number}{8}
  (\bibinfo{year}{2019}), \bibinfo{pages}{1881--1897}.
\newblock


\end{thebibliography}

\end{document}